\documentclass{elsart}
\usepackage{graphicx,amsmath}
\newcounter{bla}
\newenvironment{refnummer}{%
\list{[\arabic{bla}]}%
{\usecounter{bla}%
 \setlength{\itemindent}{0pt}%
 \setlength{\topsep}{0pt}%
 \setlength{\itemsep}{0pt}%
 \setlength{\labelsep}{2pt}%
 \setlength{\listparindent}{0pt}%
 \settowidth{\labelwidth}{[9]}%
 \setlength{\leftmargin}{\labelwidth}%
 \addtolength{\leftmargin}{\labelsep}%
 \setlength{\rightmargin}{0pt}}}
 {\endlist}

\begin{document}

%%%%%%%%%%%%%%%%%%%%%%%% NEW DEFINITIONS
\def\la{\mathrel{\mathpalette\fun <}}
\def\ga{\mathrel{\mathpalette\fun >}}
\def\fun#1#2{\lower3.6pt\vbox{\baselineskip0pt\lineskip.9pt
\ialign{$\mathsurround=0pt#1\hfil##\hfil$\crcr#2\crcr\sim\crcr}}} 
\newcommand{\Pom}{{\hspace{-0.1em}I\hspace{-0.25em}P}}

\begin{frontmatter}

\title{Heavy ion event generator HYDJET++ (HYDrodynamics plus JETs)}

\author{I.P. Lokhtin\thanksref{author}},
\author{L.V. Malinina, S.V. Petrushanko, A.M. Snigirev}

\thanks[author]{Corresponding author}

\address{M.V. Lomonosov Moscow State University, D.V. Skobeltsyn
Institute of Nuclear Physics, Moscow, Russia}

\author{I. Arsene\thanksref{onleave}}, 
\author{K. Tywoniuk\thanksref{also}}

\address{The Department of Physics, University of Oslo, Norway}

\thanks[onleave]{On leave from the Institute for Space Sciences, Bucharest, Romania}
\thanks[also]{Curren affiliation: Departamento de F{\'\i}sica de
Part{\'\i}culas, Universidad de Santiago de Compostela, Santiago de Compostela, 
Espana}

\begin{abstract}
  %Type your abstract here.

HYDJET++ is a Monte-Carlo event generator for simulation of relativistic heavy 
ion AA collisions considered as a superposition of the soft, hydro-type state and 
the hard state resulting from multi-parton fragmentation. This model is the development 
and continuation of HYDJET event generator (Lokhtin \& Snigirev, 2006, EPJC, 45, 211). 
The main program is written in the object-oriented C++ language under the ROOT environment. 
The hard part of HYDJET++ is identical to the hard part of Fortran-written HYDJET 
and it is included in the generator structure as a separate directory. The soft part 
of HYDJET++ event is the ``thermal'' hadronic state generated on the chemical and 
thermal freeze-out hypersurfaces obtained from the parameterization of relativistic 
hydrodynamics with preset freeze-out conditions. It includes the longitudinal, radial 
and elliptic flow effects and the decays of hadronic resonances. The corresponding fast 
Monte-Carlo simulation procedure, C++ code FAST MC (Amelin et al., 2006, PRC, 74, 
064901; 2008, PRC, 77, 014903) is adapted to HYDJET++. It is designed for studying the 
multi-particle production in a wide energy range of heavy ion experimental facilities: 
from FAIR and NICA to RHIC and LHC. 

\medskip

\begin{flushleft}
  %Insert your suggested PACS number here
PACS: 24.10.Lx, 24.85.+p, 25.75.-q, 25.75.Bh, 25.75.Dw, 25.75.Ld, 25.75.Nq 	
\end{flushleft}

\begin{keyword}
Monte-Carlo event generators, relativistic heavy ion collisions, hydrodynamics,
QCD jets, partonic energy loss, flow, quark-gluon plasma

  % Please give some freely chosen keywords that we can use in a
  % cumulative keyword index.
\end{keyword}

\end{abstract}
 
\end{frontmatter}

% Computer program descriptions should contain the following
% PROGRAM SUMMARY.

\newpage

{\bf PROGRAM SUMMARY}
  %Delete as appropriate.

\begin{small}
\noindent
{\em Manuscript Title:} Heavy ion event generator HYDJET++ (HYDrodynamics plus JETs) 

\noindent
{\em Authors:} Igor Lokhtin, Ludmila Malinina, Sergey Petrushanko, 
Alexander Snigirev, Ionut Arsene, Konrad Tywoniuk 

\noindent
{\em Program Title:} HYDJET++, version 2.0                    

%{\em Journal Reference:}                                      \\
  %Leave blank, supplied by Elsevier.
%{\em Catalogue identifier:}                                   \\
  %Leave blank, supplied by Elsevier.
%{\em Licensing provisions:} none                                  \\
  %enter "none" if CPC non-profit use license is sufficient.

\noindent
{\em Programming language:} C++ (however there is a Fortran-written part which is 
included in the generator structure as a separate directory)

\noindent
{\em Size of the package:} 3.5 MBytes directory and 800 kBytes compressed
distribution archive (without ROOT libraries). The output file created by
the code in ROOT tree format for 100 central (0$-$5\%) Au+Au events at 
$\sqrt{s} = 200 A$ GeV (Pb+Pb events at $\sqrt{s} = 5500 A$ GeV) with default 
input parameters requires $40$ ($190$) MBytes of the disk space.  

\noindent
{\em Computer:} PC - hardware independent 
(both C++ and Fortran compilers and ROOT environment \cite{root} should be 
installed)
  %Computer(s) for which program has been designed.

\noindent  
{\em Operating system:} Linux (Scientific Linux, Red Hat Enterprise, FEDORA, 
etc.)
  %Operating system(s) for which program has been designed.

\noindent
{\em RAM:} 50 MBytes (determined by ROOT requirements)     
  %RAM in bytes required to execute program with typical data.

\noindent
{\em Number of processors used:} 1                         
  %If more than one processor.
      
\noindent    
{\em Keywords:} Monte-Carlo event generators, relativistic heavy ion collisions, 
hydrodynamics, QCD jets, partonic energy loss, flow, quark-gluon plasma  
  % Please give some freely chosen keywords that we can use in a
  % cumulative keyword index.
  
\noindent  
{\em PACS:} 24.10.Lx, 24.85.+p, 25.75.-q, 25.75.Bh, 25.75.Dw, 25.75.Ld, 25.75.Nq 
  % see http://www.aip.org/pacs/pacs.html
  
\noindent  
{\em Classification:} 11.2 Elementary particle physics, phase space and event
simulation                                  
  %Classify using CPC Program Library Subject Index, see (
  % http://cpc.cs.qub.ac.uk/subjectIndex/SUBJECT_index.html)
  %e.g. 4.4 Feynman diagrams, 5 Computer Algebra.
  
\noindent  
{\em External routines/libraries:} ROOT (any version)
  % Fill in if necessary, otherwise leave out.
  
\noindent  
{\em Subprograms used:} PYTHIA event generator (version 6.401 or later), 
PYQUEN event generator (version 1.5 or later)  
  %Fill in if necessary, otherwise leave out.

%{\em Catalogue identifier of previous version:}*              \\
  %Only required for a New Version summary, otherwise leave out.
%{\em Journal reference of previous version:}*                  \\
  %Only required for a New Version summary, otherwise leave out.
%{\em Does the new version supersede the previous version?:}*   \\
  %Only required for a New Version summary, otherwise leave out.

\noindent
{\em Nature of the physical problem:}\\
  %Describe the nature of the problem here.
The experimental and phenomenological study of multi-particle production 
in relativistic heavy ion collisions is expected to provide 
valuable information on the dynamical behaviour of strongly-interacting matter 
in the form of quark-gluon plasma (QGP)~\cite{qm05,qm06,qm08}, as predicted by 
lattice Quantum Chromodynamics (QCD) calculations. Ongoing and future  
experimental studies in a wide range of heavy ion beam energies require the 
development of new Monte-Carlo (MC) event generators  
and improvement of existing ones. Especially for experiments 
at the CERN Large Hadron Collider (LHC), implying very high parton 
and hadron multiplicities, one needs fast (but realistic) MC tools for heavy 
ion event simulations~\cite{alice1,alice2,cms}. The main advantage of 
MC technique for the simulation of high-multiplicity hadroproduction is that it allows  
a visual comparison of theory and data, including if necessary the 
detailed detector acceptances, responses and resolutions. The realistic MC 
event generator has to include maximum possible number of observable physical 
effects, which are important to determine the event topology: from the 
bulk properties of soft hadroproduction (domain of low transverse momenta 
$p_T \la 1$GeV$/c$) such as collective flows, to hard multi-parton production 
in hot and dense QCD-matter, which reveals itself in the spectra of high-$p_T$ 
particles and hadronic jets. Moreover, the role of hard and semi-hard particle 
production at LHC can be significant even for the bulk properties of created matter, 
and hard probes of QGP became clearly observable in various new 
channels~\cite{Abreu:2007kv,hp1,hp2,hp3}. In majority of the available MC 
heavy ion event generators, the simultaneous treatment of collective flow 
effects for soft hadroproduction and hard multi-parton in-medium production 
(medium-induced partonic rescattering and energy loss, so called ``jet
quenching'') is lacking. Thus, in order to analyze existing data on low and 
high-p$_T$ hadron production, test the sensitivity of physical observables 
at the upcoming LHC experiments (and other future heavy ion facilities) to the QGP 
formation, and study the experimental capabilities of constructed detectors, the 
development of adequate and fast MC models for simultaneous collective 
flow and jet quenching simulations is necessary. HYDJET++ event generator 
includes detailed treatment of soft hadroproduction as well as hard 
multi-parton production, and takes into account known medium effects. 

\noindent   
{\em Solution method:}\\
  %Describe the method solution here.
 
A heavy ion event in HYDJET++ is a superposition of the 
soft, hydro-type state and the hard state resulting from multi-parton fragmentation. 
Both states are 
treated independently. HYDJET++ is the development and continuation of HYDJET 
MC model~\cite{Lokhtin:2005px}. The main program is written in 
the object-oriented C++ language under the ROOT environment~\cite{root}. The 
hard part of HYDJET++ is identical to the hard part of 
Fortran-written HYDJET~\cite{hydjet} (version 1.5) and is included in the generator 
structure as a separate directory. The routine for generation of single hard NN 
collision, generator PYQUEN \cite{Lokhtin:2005px,pyquen}, modifies  
the ``standard'' jet event obtained with the generator PYTHIA$\_$6.4~\cite{pythia}. 
The event-by-event simulation procedure 
in PYQUEN includes {\it 1)} generation of initial parton spectra with PYTHIA and 
production vertexes at given impact parameter; {\it 2)} rescattering-by-rescattering 
simulation of the parton path in a dense zone and its radiative and collisional 
energy loss; {\it 3)} final hadronization according to the Lund string model for hard 
partons and in-medium emitted gluons. Then the PYQUEN multi-jets generated 
according to the binomial distribution are included in the hard part of the event. 
The mean number of jets produced in an AA event is the  
product of the number of binary NN sub-collisions at a given impact parameter   
and the integral cross section of the hard process in $NN$ collisions with the minimum  
transverse momentum transfer $p_T^{\rm min}$. In order to take into account the 
effect of nuclear shadowing on parton distribution functions, the impact parameter 
dependent parameterization obtained in the framework of Glauber-Gribov 
theory~\cite{Tywoniuk:2007xy} is used. The soft part of HYDJET++ 
event is the ``thermal'' hadronic state generated on the chemical and thermal 
freeze-out hypersurfaces obtained from the parameterization of relativistic 
hydrodynamics with preset freeze-out conditions (the adapted C++ code 
FAST MC~\cite{Amelin:2006qe,Amelin:2007ic}). 
Hadron multiplicities are calculated using the effective thermal volume 
approximation and Poisson multiplicity distribution around its mean value, 
which is supposed to be proportional to the number of participating nucleons at 
a given impact parameter of AA collision. The fast soft hadron simulation procedure 
includes {\it 1)} generation of the 4-momentum of a hadron in the rest frame of a 
liquid element in accordance with the equilibrium distribution function; {\it 2)}  
generation of the spatial position of a liquid element and its local 4-velocity in 
accordance with phase space and the character of motion of the fluid; {\it 3)} the
standard von Neumann rejection/acceptance procedure to account for the difference
between the true and generated probabilities; {\it 4)} boost of the hadron 4-momentum in the center mass frame of the event; {\it 5)} the two- and three-body decays 
of resonances with branching ratios taken from the SHARE particle decay 
table~\cite{share}. The high generation speed in HYDJET++ is achieved due to  
almost 100\% generation efficiency of the ``soft'' part because of the  
nearly uniform residual invariant weights which appear in the freeze-out 
momentum and coordinate simulation. Although HYDJET++ is optimized for very 
high energies of RHIC and LHC colliders (c.m.s. energies of 
heavy ion beams $\sqrt{s}=200$ and $5500$ GeV per nucleon pair 
respectively), in practice it can also be used for studying the  
particle production in a wider energy range down to $\sqrt{s} \sim 10$ GeV per nucleon pair at other heavy ion experimental facilities. As one moves from 
very high to moderately high energies, the contribution of the hard part of the event 
becomes smaller, while the soft part turns into just a multi-parameter fit to the data.
   
%{\em Reasons for the new version:}*\\
  %Only required for a New Version summary, otherwise leave out.
%   \\
%{\em Summary of revisions:}*\\
  %Only required for a New Version summary, otherwise leave out.
%   \\

\noindent
{\em Restrictions:}\\
  %Describe any restrictions on the complexity of the problem here.
  HYDJET++ is only applicable for symmetric AA collisions of heavy (A $\ga 40$) ions 
  at high energies (c.m.s. energy $\sqrt{s} \ga 10$ GeV per nucleon pair). The 
  results obtained for very peripheral collisions (with the impact parameter of the order 
  of two nucleus radii, $b \sim 2 R_A$) and very forward rapidities may be not 
  adequate. 

%\noindent   
%{\em Unusual features:}\\
  %Describe any unusual features of the program/problem here.

%\noindent   
%{\em Additional comments:}\\
  %Provide any additional comments here.

\noindent   
{\em Running time:}\\
The generation of 100 central (0$-$5\%) Au+Au events at $\sqrt{s} = 200 A$ GeV 
(Pb+Pb events at $\sqrt{s} = 5500 A$ GeV) with default input parameters 
takes about 7 (85) minutes on a PC 64 bit Intel Core Duo CPU @ 3 GHz with 8 
GB of RAM memory under Red Hat Enterprise.
  %Give an indication of the typical running time here.

\noindent   
{\em Accessibility:} \\
http://cern.ch/lokhtin/hydjet++

\noindent   
{\em References for the physics model:}
\begin{refnummer}
\item  I.P. Lokhtin, A.M. Snigirev, Eur. Phys. J. C 46 (2006) 211.
\item N.S. Amelin, R. Lednicky, T.A. Pocheptsov, I.P. Lokhtin, L.V. Malinina, 
A.M. Snigirev, Iu.A. Karpenko and Yu.M. Sinyukov, Phys. Rev. C 74 (2006) 064901.
\item N.S. Amelin, I. Arsene, L. Bravina, Iu.A. Karpenko, R. Lednicky, I.P. 
Lokhtin, L.V. Malinina, A.M. Snigirev and Yu.M. Sinyukov, Phys. Rev. C 77 
(2008) 014903.
\end{refnummer}
\end{small}

\newpage

% In program descriptions the main text of the paper is listed under
% the heading LONG WRITE-UP.

\section{Introduction}

One of the basic tasks of modern high energy physics is the study of the 
fundamental theory of strong interaction (Quantum Chromodynamics, QCD) in 
new, unexplored extreme regimes of super-high densities and temperatures through  
the investigation of the properties of hot and dense multi-parton and multi-hadron systems produced in high-energy nuclear 
collisions~\cite{Hwa:2004yg,d'Enterria:2006su,BraunMunzinger:2007zz}. Indeed, 
QCD is not just a quantum field theory with an extremely rich dynamical content 
(such as asymptotic freedom, chiral symmetry, non-trivial vacuum topology, 
strong CP violation problem, colour superconductivity), but perhaps the only 
sector of the Standard Model, where the basic features (as phase diagram, 
phase transitions, thermalisation of fundamental fields) may be the subject 
to scrutiny in the laboratory. The experimental and phenomenological study of 
multi-particle production in ultrarelativistic heavy ion 
collisions~\cite{qm05,qm06,qm08} is expected to provide valuable information on 
the dynamical behaviour of QCD matter in the form of a quark-gluon plasma 
(QGP), as predicted by lattice calculations. 

Experimental data, obtained from the Relativistic Heavy Ion Collider (RHIC) at 
maximum beam energy in the center of mass system of colliding ions $\sqrt{s}=200$ GeV 
per nucleon pair, supports the picture of formation of a strongly interacting 
hot QCD matter (``quark-gluon fluid'') in the most central Au+Au (and likely Cu+Cu) 
collisions~\cite{brahms05,phobos05,star05,phenix05}. This appears as    
significant modification of properties of multi-particle production in heavy
ion collisions as compared with the corresponding proton-proton (or peripheral
heavy ion) interactions. In particular, one of the important perturbative 
(``hard'') probes of QGP is the medium-induced energy loss of energetic 
partons, so called ``jet quenching''~\cite{Baier:2000m}, which is predicted to 
be very different in cold nuclear matter and in QGP, and leads to a number of 
phenomena which are already seen in the RHIC data on the qualitative level, such
as suppression of high-$p_T$ hadron production, 
modification of azimuthal back-to-back correlations, azimuthal anisotropy of 
hadron spectra at high p$_T$, etc.~\cite{Wang:2003aw}. On the other hand, one of 
the most spectacular features of low transverse momentum (``soft'') 
hadroproduction at RHIC are strong collective flow effects: the radial flow 
(ordering of the mean transverse momentum of hadron species with corresponding  
mass) and the elliptic flow (mass-ordered azimuthal anisotropy of particle yields 
with respect to the reaction plane in non-central collisions). The development 
of such a strong flow is well described by the hydrodynamic models and 
requires short time scale and large pressure gradients, attributed to strongly 
interacting systems~\cite{Heinz:2004ar}. Note however, that results of 
hydrodynamic models usually disagree with the data on femtoscopic momentum 
correlations, resulting from the space-time characteristics of the system at 
freeze-out stage.

The heavy ion collision energy in Large Hadron Collider (LHC) at CERN  
a factor of $~30$ larger then that in RHIC, thereby allows one to probe new 
frontiers of super-high temperature and (almost) net-baryon free 
QCD~\cite{Abreu:2007kv,hp1,hp2,hp3}. The emphasis of the LHC heavy ion data 
analysis (at $\sqrt{s}=5.5$ TeV per nucleon pair for lead beams) will be on the 
perturbative, or hard probes of the QGP (quarkonia, jets, photons, high-p$_T$ 
hadrons) as well as on the global event properties, or soft probes (collective 
radial and elliptic flow effects, hadron multiplicity, transverse energy 
densities and femtoscopic momentum correlations). It is expected 
that at LHC energies the role of hard and semi-hard particle production will be 
significant even for the bulk properties of created matter. 

Another domain of QCD phase diagram is high baryon density region, which can be 
probed in the future experimental studies at relatively moderate heavy ion beam
energies $\sqrt{s} \sim 10$ GeV per nucleon pair (projects CBM~\cite{cbm} and 
MPD~\cite{nica} at accelerator facilities FAIR-GSI and NICA-JINR, 
programs for heavy ion runs with lower beam energies at SPS and RHIC). These 
studies are motivated by the intention to search for the ``critical point'' of the
quark-hadron phase transition, predicted by lattice QCD, where the type of  
transition is changing from smooth ``crossover'' (at high temperatures and low 
net-baryon densities) to first order transition (at high net-baryon densities and low 
temperatures)~\cite{cp}.  

Ongoing and future experimental studies of relativistic heavy ion collisions 
in a wide range of beam energies require the development of new Monte-Carlo 
(MC) event generators and improvement of existing ones. Especially for 
experiments which will be conducted at LHC, due to very high parton 
and hadron multiplicities, one needs fast (but realistic) MC tools for heavy 
ion event simulation~\cite{alice1,alice2,cms}. The main advantage of MC technique 
for the simulation of high-multiplicity hadroproduction is that it allows a visual  
comparison of theory and data, including if necessary the detailed detector 
acceptances, responses and resolutions. A realistic MC event generator should  
include a maximum possible number of observable physical effects which are 
important to determine the event topology: from the bulk properties of 
soft hadroproduction (domain of low transverse momenta $p_T \la 1$GeV$/c$)  
such as collective flows, to hard multi-parton production in hot and dense 
QCD-matter, which reveals itself in the spectra of high-$p_T$ particles and 
hadronic jets. 

In most of the available MC heavy ion event generators, the 
simultaneous treatment of collective flow effects for soft hadroproduction and 
hard multi-parton in-medium production is absent. For example, the popular
MC model HIJING~\cite{hijing} includes jet production and jet quenching 
on some level, but it does not include any significant flow effects. The recently 
developed MC model JEWEL~\cite{Zapp:2008gi}, which combines a 
perturbative final state parton shower with QCD-medium effects, is actually not a 
heavy ion event generator, but rather a simulator of jet quenching in individual 
nucleon-nucleon sub-collisions (like PYQUEN~\cite{Lokhtin:2005px,pyquen}). 
The event generators FRITIOF~\cite{fritiof} and LUCIAE~\cite{luciae} 
include jet production (but without jet quenching), while some collective nuclear 
effects (such as string interactions and hadron rescatterings) are taken into 
account in LUCIAE. Another MC model 
THERMINATOR~\cite{therminator} includes detailed statistical description of 
``thermal'' soft particle production and can reproduce the main bulk features 
of hadron spectra at RHIC (in particular, describe simultaneously the 
momentum-space measurements and the freeze-out coordinate-space data), but it 
does not include hard parton production processes. There is a number of 
microscopic transport hadron models (UrQMD~\cite{urqmd}, QGSM~\cite{qgsm}, 
AMPT~\cite{ampt}, etc.), which attempt to analyze the soft particle 
production in a wide energy range, however they also do not include in-medium 
production of high-$p_T$ multi-parton states. Moreover, hadronic cascade 
models have difficulties with the detailed description of relatively 
low-$p_T$ RHIC data on elliptic flow and the size of hadron emission, obtained from 
momentum correlation measurements (e.g. AMPT can reproduce the elliptic flow or  
the correlation radii using different sets of model parameters). Another  
heavy ion event generator, ZPC~\cite{zpc}, has been created to simulate parton 
cascade evolution in ultrarelativistic heavy ion collisions. From a physical point 
of view, such an approach seems reasonable only for very high beam energies (RHIC, 
LHC). Note also that the full treatment of parton cascades requires enormous 
amount of CPU run time (especially for LHC).

Thus, in order to analyze existing data on low and high-p$_T$ hadron production, 
test the sensitivity of physical observables at the upcoming LHC experiments 
(and other future heavy ion facilities) to the QCD-matter formation, and study 
the experimental capabilities of constructed detectors, the development of 
adequate and fast MC models for simultaneous collective flow and jet 
quenching simulation is necessary. HYDJET++ event generator includes detailed 
treatment of soft hadroproduction as well as hard multi-parton production, and 
takes into account medium-induced parton rescattering and energy loss. Although 
the model is optimized for very high energies of colliding nuclei (RHIC, LHC), 
in practice it can also be used for studying the multi-particle production at  
lower energies at other heavy ion experimental facilities as FAIR and NICA. 
Moving from very high to moderately high energies, the contribution of the hard part 
of the event becames smaller, while the soft part turns into just a multi-parameter 
fit to the data. The heavy ion event in HYDJET++ is the 
superposition of two independent parts: the soft, hydro-type state and the hard state
resulting from  
multi-parton fragmentation. Note that a conceptually similar approximation has been 
developed in~\cite{Hirano:2004rs} but, to our best knowledge, it is not implemented 
as an MC event generator. 

The main program of HYDJET++ is written in the object-oriented C++ language under the ROOT environment~\cite{root}. The hard part of HYDJET++ is identical to the hard part of 
Fortran-written HYDJET (version 1.5) and it is included in the generator structure as 
a separate directory. The soft part of HYDJET++ event is the ``thermal'' hadronic 
state generated on the chemical and thermal freeze-out hypersurfaces obtained  
from a parameterization of relativistic hydrodynamics with preset freeze-out 
conditions. It includes the longitudinal, radial and elliptic flow effects and the decays of hadronic resonances. The corresponding fast MC simulation procedure based on 
C++ code, FAST MC~\cite{Amelin:2006qe,Amelin:2007ic}, was adapted to HYDJET++. The 
high generation speed in FAST MC is achieved due to almost 100\% generation 
efficiency because of nearly uniform residual invariant weights which  
appear in the freeze-out momentum and coordinate simulation.
  
Let us indicate some physical restrictions of the model. HYDJET++ is only 
applicable for symmetric AA collisions of heavy (A $\ga 40$) ions at high 
energies ($\sqrt{s} \ga 10$ GeV). Since the hydro-type approximation for heavy 
ion collisions is considered to be valid for central and semi-central collisions, 
the results obtained for very peripheral collisions (with impact parameter of 
the order of two nucleus radii, $b \sim 2 R_A$) may be not adequate. Nor do we  
expect a correct event description in the region of very forward rapidities, where the other mechanisms of particle production, apart from hydro-flow and jets, may be important. 

\section{Physics model}

A heavy ion event in HYDJET++ is a superposition of the soft, hydro-type state  
and the hard state resulting from multi-parton fragmentation. Both states are treated 
independently. 

\subsection{Model for the hard multi-jet production}

The model for the hard multi-parton part of HYDJET++ event is the same as that 
for HYDJET event generator. A detailed description of the physics framework of 
this model can be found in the corresponding paper~\cite{Lokhtin:2005px}. The 
approach to the description of multiple scattering of hard partons in the dense 
QCD-matter (such as quark-gluon plasma) is based on the accumulative energy loss via  the gluon radiation being associated with each parton scattering in the expanding quark-gluon fluid and includes the interference effect 
(QCD analog~\cite{Wang:1994fx,Baier:1996kr,Zakharov:1997uu,Gyulassy:2000fs,
Wiedemann:2000tf} of the well known Landau-Pomeranchuk-Migdal (LPM) effect in 
QED~\cite{lpm1,lpm2} for the emission of gluons 
with a finite formation time) using the modified radiation spectrum $dE/dl$ as a 
function of decreasing temperature $T$. The basic kinetic integral equation for 
the energy loss $\Delta E$ as a function of initial energy $E$ and path length $L$ 
has the form 
\begin{eqnarray} 
\label{elos_kin}
\Delta E (L,E) = \int\limits_0^Ldl\frac{dP(l)}{dl}
\lambda(l)\frac{dE(l,E)}{dl} \, , ~~~~ 
\frac{dP(l)}{dl} = \frac{1}{\lambda(l)}\exp{\left( -l/\lambda(l)\right) }
\, ,  
\end{eqnarray} 
where $l$ is the current transverse coordinate of a parton, $dP/dl$ is the 
scattering probability density, $dE/dl$ is the energy loss per unit length, 
$\lambda = 1/(\sigma \rho)$ is the in-medium mean free path, $\rho \propto T^3$ is 
the medium density at the temperature $T$, $\sigma$ is the integral cross 
section for the parton interaction in the medium. Both collisional and radiative
energy loss are taken into account in the model. 

The partonic collisional energy loss due to elastic scatterings is treated in 
high-momentum transfer limit~\cite{bjork82,braaten91,Lokhtin:2000wm}:  
\begin{eqnarray}
\label{col} 
\frac{dE}{dl}^{col} = \frac{1}{4T \lambda \sigma} 
\int\limits_{\displaystyle \mu^2_D}^
{\displaystyle t_{\rm max}}dt\frac{d\sigma }{dt}t \, ,
\end{eqnarray} 
where the dominant contribution to the differential scattering cross section is 
\begin{eqnarray} 
\label{sigt} 
\frac{d\sigma }{dt} \cong C \frac{2\pi\alpha_s^2(t)}{t^2} 
\frac{E^2}{E^2-m_p^2} \, , ~~~~ 
\alpha_s = \frac{12\pi}{(33-2N_f)\ln{(t/\Lambda_{QCD}^2)}} \> \,
\end{eqnarray} 
for the scattering of a hard parton with energy $E$ and mass $m_p$ off the ``thermal'' 
parton with energy (or effective mass) $m_0 \sim 3T \ll E$. Here $C = 9/4, 1, 4/9$ 
for $gg$, $gq$ and $qq$ scatterings respectively, $\alpha_s$ is the QCD running 
coupling constant for $N_f$ active quark flavors, and $\Lambda_{QCD}$ is the QCD 
scale parameter which is of the order of the critical temperature of quark-hadron 
phase transition,  $\Lambda_{QCD}\simeq T_c \simeq 200$ MeV. The integrated cross 
section $\sigma$ is regularized by the Debye screening mass squared 
$\mu_D^2 (T) \simeq 4\pi \alpha _s T^2(1+N_f/6)$. The maximum momentum transfer 
$t_{\rm max}=[ s-(m_p+m_0)^2] [ s-(m_p-m_0)^2 ] / s$ where $s=2m_0E+m_0^2+m_p^2$.
The model simplification is that the collisional energy loss due to scatterings 
with low momentum transfer (resulting mainly from the interactions with plasma 
collective modes or colour background fields, see for recent 
developments~\cite{Randrup:2003cw,Markov:2003rk,Mustafa:2003vh,Peigne:2005rk,
Zapp:2005kt,Adil:2006ei,Djordjevic:2006tw,Alam:2006qf,Ayala:2007cq} and references therein) is not considered here. Note, however, that in the majority of estimations,  the latter process does not contribute much to the total collisional loss 
in comparison with the high-momentum scattering (due to absence of the large factor 
$\sim \ln{(E / \mu_D)}$), and in numerical computations it can be effectively 
``absorbed'' by means of redefinition of minimum momentum transfer $t_{\rm min} 
\sim \mu_D^2$ in (\ref{col}). Another model assumption is that the collisional loss
represents the incoherent sum over all scatterings, although it was argued recently 
in~\cite{Wang:2006qr} that interference effects may appear in elastic parton 
energy loss in a medium of finite size. 

The partonic radiative energy loss is treated in the frameworks of BDMS 
formalism~\cite{Baier:1999ds,Baier:2001qw}. In fact, there are several 
calculations of the inclusive energy distribution of medium-induced gluon 
radiation using Feyman multiple scattering diagrams. The detailed discussions 
on the relation between these approaches, their basic parameters and physics 
predictions can be found in the recent proceedings of CERN TH Workshops (see 
~\cite{Abreu:2007kv,hp1} and references therein). In the BDMS approach, 
the strength of multiple scattering is characterized by the transport coefficient 
$\hat{q}=\mu_D^2/\lambda_g $ ($\lambda_g$ is the gluon mean free path), which is
related to the elastic scattering cross section $\sigma$ in (\ref{sigt}). In our
simulations this strength is regulated mainly by the initial QGP 
temperature $T_0$. Then the energy spectrum of coherent medium-induced gluon 
radiation and the corresponding dominant part of radiative energy loss of
massless parton become~\cite{Baier:1999ds,Baier:2001qw}: 
\begin{eqnarray} 
\label{radiat} 
\frac{dE}{dl}^{rad} = \frac{2 \alpha_s (\mu_D^2) C_R}{\pi L}
\int\limits_{\omega_{\min}}^E  
d \omega \left[ 1 - y + \frac{y^2}{2} \right] 
\>\ln{\left| \cos{(\omega_1\tau_1)} \right|} 
\>, \\  
\omega_1 = \sqrt{i \left( 1 - y + \frac{C_R}{3}y^2 \right)   
\bar{\kappa}\ln{\frac{16}{\bar{\kappa}}}}
\quad \mbox{with}\quad 
\bar{\kappa} = \frac{\mu_D^2\lambda_g  }{\omega(1-y)} ~, 
\end{eqnarray} 
where $\tau_1=L/(2\lambda_g)$, $y=\omega/E$ is the fraction of the hard parton 
energy carried away by the radiated gluon, and $C_R = 4/3$ is the quark color 
factor. A similar expression for the gluon jet can be obtained by setting  
$C_R=3$ and properly changing the factor in the square brackets in 
(\ref{radiat})~\cite{Baier:1999ds}. The integration (\ref{radiat}) is carried out 
over all energies from $\omega_{\min}=E_{\rm LPM}=\mu_D^2\lambda_g$, the minimum 
radiated gluon energy in the coherent LPM regime, up to initial parton energy $E$. 
Note that we do not consider here possible effects of double parton 
scattering~\cite{Wang:2001ifa,Vitev:2005yg} and thermal gluon 
absorption~\cite{Wang:2001cs}, which can be included in the model in the future.  

The simplest generalization of the formula (\ref{radiat}) for a heavy quark of mass 
$m_q$ can be done by using the ``dead-cone'' approximation~\cite{Dokshitzer:2001zm}:  
\begin{equation}
\label{radmass} 
\frac{dE}{dld\omega }| _{m_q \ne 0} =  \frac{1}{(1+(\beta \omega )^{3/2})^2}
\frac{dE}{dld\omega }| _{m_q=0}, ~~~ \beta =\left( \frac{\lambda}{\mu_D^2}\right) 
^{1/3} \left( \frac{m_q}{E}\right) ^{4/3}~. 
\end{equation}
One should mention a number of more recent developments in heavy quark energy loss 
calculations available in the literature (see~\cite{Abreu:2007kv} and references 
therein), which can be included in the model in the future.

The medium where partonic rescattering occurs is treated as a boost-invariant 
longitudinally expanding quark-gluon fluid, and the partons as being produced on a 
hyper-surface of equal proper times $\tau$~\cite{bjork86}. In order to simplify 
numerical calculations we omit here the transverse expansion and viscosity of the 
fluid using the well-known scaling solution obtained by Bjorken~\cite{bjork86} for 
a temperature $T$ and energy density $\varepsilon$ of QGP at $T > T_c \simeq 200$ 
MeV:
\begin{equation}
\varepsilon(\tau) \tau^{4/3} = \varepsilon_0 \tau_0^{4/3},~~~~
T(\tau) \tau^{1/3} = T_0 \tau_0^{1/3},~~~~ \rho(\tau) \tau = \rho_0 \tau_0 .
\end{equation}
The proper time $\tau_0$ of QGP formation, the initial QGP temperature $T_0$ at 
mid-rapidity ($y=0$) for central (impact parameter $b=0$) Pb+Pb collisions, and 
the number $N_f$ of active flavours are input model parameters. For non-central 
collisions and for other beam atomic numbers, $T_0$ is calculated automatically:
the initial energy density $\varepsilon _0 (b)$ is supposed to be proportional to 
the ratio of nuclear overlap function $T_{AA}(b)$ to effective transverse area 
of nuclear overlapping $S_{AA}(b)$~\cite{Lokhtin:2000wm}: 
\begin{eqnarray} 
\label{eps_0} 
& & \varepsilon_0(b)=\varepsilon_0(b=0)\frac{T_{AA}(b)}{T_{AA}(b=0)} 
\frac{S_{AA}(b=0)}{S_{AA}(b)} ~,~\varepsilon_0(A) =
\varepsilon_0(Pb)\left( \frac{A}{207}\right) ^{2/3}, \\
\label{taa}
& &  T_{AA}(b)=\int\limits_0^{2\pi} d \psi \int\limits_0^{\infty }rdrT_A(r_1)
T_A(r_2)~,
~~~~S_{AA}(b)=\int\limits_0^{2\pi} d\psi \int\limits_0^{R_{\rm eff}(b, \psi)}rdr~,\\
\label{ta} 
& & T_A({\bf r}) = A \int \rho_A({\bf r},z)dz~,
~~~~~~~~r_{1,2} = \sqrt{r^2+\frac{b^2}{4}\pm rb\cos \psi}~,\\
\label{eff} 
& & R_{\rm eff} = \min \{ \sqrt{R_A^2 - \frac{b^2}{4} \sin^2 \psi} + \frac{b}{2} 
\cos \psi ,  \sqrt{R_A^2 - \frac{b^2}{4} \sin^2 \psi} - \frac{b}{2} \cos \psi \}, 
\end{eqnarray} 
where $r_{1,2} (b,r,\psi)$ are the distances between the centres of colliding nuclei and the 
jet production vertex $V(r\cos{\psi}, r\sin{\psi})$ ($r$ being the distance from 
the nuclear collision axis to V), $R_{\rm eff}(b, \psi)$ is the transverse distance 
from the nuclear collision axis to the effective boundary of nuclear overlapping area 
in the given azimuthal direction $\psi$ (so that $R_{\rm eff}(b=0)$ is equal to the 
nuclear radius $R_A=1.15 A^{1/3}$), $T_A(b)$ is the nuclear thickness functions, and 
$\rho_A({\bf r},z)$ is the standard Woods-Saxon nucleon density distribution~\cite{ws} 
(see \cite{Lokhtin:2000wm} for detailed nuclear geometry explanations). The 
rapidity dependent spreading of the initial energy density around mid-rapidity 
$y=0$ is taken in the Gaussian-like form. The radial profile of the energy density 
(or the temperature) is also introduced in the parton rescattering model. The 
transverse energy density in each point inside the nuclear overlapping zone is 
proportional to the impact-parameter dependent product of two nuclear thickness 
functions $T_A$ in this point: 
\begin{eqnarray}
\label{enprof} 
& & \varepsilon (r_1, r_2) \propto T_A(r_1) T_A(r_2)~.  
\end{eqnarray}
Thus, in fact, the input parameter $T_0 \propto \varepsilon_0^{1/4}$ acquires the 
meaning of an effective (i.e. averaged over the whole nuclear overlapping region) 
initial temperature in central Pb+Pb collisions. 

Note that the other scenarios of QGP space-time evolution for the MC 
implementation of the model were also studied. In particular, the influence of 
the transverse flow, as well as that of the mixed phase at $T = T_c$, on the 
intensity of jet rescattering (which is a strongly increasing function of $T$) 
has been found to be inessential for high initial temperatures $T_0 \gg T_c$. 
On the contrary, the presence of QGP viscosity slows down the cooling rate, 
that implies a jet parton spending more time in the hottest regions of the 
medium. As a result the rescattering intensity increases, i.e., in fact the  
effective temperature of the medium appears to be higher as compared with the perfect QGP case. We also do not take into account here the probability of jet rescattering 
in the nuclear matter, because the intensity of this process and the corresponding 
contribution to the total energy loss are not significant due to much smaller 
energy density in a ``cold'' nucleus.

Another important element of the model is the angular spectrum of in-medium gluon 
radiation. Since the detailed calculation of the angular spectrum of emitted gluons 
is rather sophisticated, the simple ``small-angle'' parameterization of the 
gluon distribution over the emission angle $\theta$ was used:
\begin{eqnarray} 
\label{sar} 
& & \frac{dN^g}{d\theta}\propto \sin{\theta} \exp{\left( -\frac{(\theta-\theta
_0)^2}{2\theta_0^2}\right) }~, 
\end{eqnarray} where $\theta_0 \sim 5^0$ is the typical angle of the coherent gluon 
radiation as estimated in~\cite{Lokhtin:1998ya}. Two other parameterizations 
(``wide-angle'' $dN^g/d\theta \propto 1/\theta$, and ``collinear''  
$dN^g/d\theta = \delta (\theta )$) are also envisaged. 

The model for single hard nucleon-nucleon sub-collision
PYQUEN~\cite{pyquen} was constructed as a modification of the jet event obtained 
with the generator of hadron-hadron interactions PYTHIA$\_$6.4~\cite{pythia}. 
The event-by-event simulation procedure in PYQUEN includes generation of 
the initial parton spectra with PYTHIA and production vertexes at given impact 
parameter, rescattering-by-rescattering simulation of the parton path in a 
dense zone, radiative and collisional energy loss per rescattering, final 
hadronization with the Lund string model for hard partons and in-medium emitted 
gluons. Then the PYQUEN multi-jets generated according to the binomial distribution 
are included in the hard part of the event. If there is no
nuclear shadowing, the mean number of jets produced in AA events at a given 
impact parameter $b$ is proportional to the mean number of binary NN sub-collisions,
$\overline{N_{\rm bin}}=T_{AA}(b) \sigma _{NN}^{\rm in}(\sqrt{s})$. In the presence of shadowing, this number is determined as
\begin{eqnarray}
\label{numjets}
\overline{N_{AA}^{{\rm jet}}} (b,\sqrt{s},p_T^{\rm min}) =    
\int\limits_{p_T^{\rm min}} dp_T^2 \int dy  
\frac{d\sigma_{NN}^{\rm hard}(p_T,~\sqrt{s})}{dp_T^2dy}
\int\limits_0^{2\pi}d \psi \int\limits_0^{\infty }rdr \nonumber \\
T_A(r_1)T_A(r_2)S(r_1,r_2,p_T,y)~, 
\end{eqnarray} 
where  $\sigma _{NN}^{\rm in}(\sqrt{s})$ and 
$d\sigma_{NN}^{\rm hard}(p_T, \sqrt{s})/dp_T^2dy$, calculated with PYTHIA, are the  
total inelastic non-diffractive NN cross section and the differential cross section 
of the corresponding hard process in $NN$ collisions (at the same c.m.s. energy, 
$\sqrt{s}$, of colliding beams) with the minimum transverse momentum transfer 
$p_T^{\rm min}$ respectively. The latter is another input parameter of the model. 
In the HYDJET frameworks, partons produced in (semi)hard processes with the momentum 
transfer lower than $p_T^{\rm min}$ are considered as being ``thermalized'', so 
their hadronization products are included in the soft part of the event 
``automatically''. The factor $S \le 1$  in (\ref{numjets}) takes into account 
the effect of nuclear shadowing on parton distribution functions. It can be 
written as a product of shadowing factors for both the colliding nuclei as
\begin{equation}
\label{S_shad}
S(r_1,r_2,p_T,y) = S^i_A(x_1,Q^2,r_1) \, S^j_A(x_2,Q^2,r_2) \;,
\end{equation}
where $S^{i,j}_A$ is the ratio of nuclear to nucleon parton distribution functions  
(normalized by the atomic number) for the parton of type $\{i,j\}$ (light quark 
or gluon), $x_{1,2}$ are the momentum fractions of the initial partons 
from the incoming nuclei which participate in the hard scattering characterized 
by the scale $Q^2 = x_1 x_2 s$, and $r_{1,2}$ are the transverse 
coordinates of the partons in their respective nuclei, so that 
$\mathbf{r_1} + \mathbf{r_2} = \mathbf{b}$. In HYDJET++, nuclear shadowing 
corrections are implemented not by modifying the parton showering of single NN 
collisions in PYTHIA but by correcting of the contribution of initial coherent, 
multiple scattering in effective way. In fact, this nuclear effect reduces 
the number of partons in the incoming hadronic wave-function of both the nuclei 
and thus reduces the total jet production cross section. Since the degree of this 
reduction depends on the kinematic variables of incoming hard partons, both initial 
and final parton momentum spectra are modified as a result of nuclear shadowing. 

The shadowing corrections are expected to be very significant at LHC energies, 
where both soft and relatively high-$p_T$ ($\la 15$ GeV/$c$) particle 
production probe the low-$x$ gluon distribution of the target at moderate 
scales, $Q^2 \sim p^2_\perp$, and are therefore strongly influenced by unitarity 
effects. In the Glauber-Gribov theory~\cite{Gribov69}, this phenomenon arises 
from coherent interaction of the projectile fluctuation on the target 
constituents and is closely related to the diffractive structure function of 
the nucleon. Due to the factorization theorem for hard processes in QCD, 
$S^i_A$ describes the modifications of nuclear parton distribution functions 
(i.e. distributions of quarks and gluons in nuclei), such that
\begin{equation}
f_{i/A}(x,Q^2,b) \;=\; f_{i/N}(x,Q^2)S_i(A,b,x,Q^2) \;.
\end{equation}
From summation of Pomeron fan diagrams the shadowing factor is found to be 
$S^i_A = 1/\left(1 + F^i(x,Q^2) T_A(b) \right)$, where the effective cross 
section for quarks and gluon, respectively, is found to be
\begin{align}
F^i(x,Q^2) =& \; 4\pi \int^{0.1}_x \!\!\!\!\!\! d x_\Pom 
\, \Pi \left( x_\Pom \right) \,
\left\{ 
\begin{array}{l}
\beta \Sigma^\mathcal{D} \left(\beta, Q^2 \right) \big/ \Sigma \left(x,Q^2 \right) \\
\beta g^\mathcal{D} \left(\beta,Q^2 \right) \big/ g\left(x,Q^2 \right)
\end{array} \right. , \\
\Pi \left( x_\Pom \right) = & \;B \left( x_\Pom \right) f_\Pom \left(x_\Pom \right) F_A^2 
\left(-x_\Pom^2 m_N^2 \right) \;,
\end{align}
where $\Sigma^\mathcal{D}$ and $g^\mathcal{D}$ denote the quark-singlet and 
gluon diffractive parton distribution functions, $\Sigma$ and $g$ are
the normal parton distribution functions, $B$ and $f_\Pom$ are the
slope of diffractive distribution and the Pomeron flux factor respectively, 
$F_A$ is the nuclear form factor. The quark and gluon diffractive 
distributions are taken from the most recent experimental parameterizations by 
the H1 Collaboration~\cite{H1diffractive}, and the resulting shadowing 
factors calculated in~\cite{Tywoniuk:2007xy} are implemented in HYDJET and 
HYDJET++.

\subsection{Model for the soft ``thermal'' hadron production}

The soft part of HYDJET++ event is the ``thermal'' hadronic state generated on 
the chemical and thermal freeze-out hypersurfaces obtained from a 
parameterization of relativistic hydrodynamics with preset freeze-out 
conditions. The detailed description of the physics frameworks of this model can be 
found in the corresponding papers~\cite{Amelin:2006qe,Amelin:2007ic}. It is
supposed that a hydrodynamic expansion of the fireball ends by a sudden system 
breakup at given temperature $T^{\rm ch}$ and chemical potentials $\mu_B, 
\mu_S, \mu_Q$ (baryon number, strangeness and electric charge respectively). In 
this case, the momentum distribution of the produced hadrons retains the thermal 
character of the (partially) equilibrated Lorentz invariant distribution 
function in the fluid element rest frame~\cite{Raf81,Gorenstein97}: 
\begin{equation}
\label{eqfun}
f_{i}^{\rm eq}(p^{*0};T^{\rm ch},\mu_i,\gamma_s) = \frac{g_i}{\gamma_s^{-n_i^s}
\exp{([p^{*0} -\mu_i]/T^{\rm ch})} \pm 1}~, 
\end{equation}
where $p^{*0}$ is the hadron energy in the fluid element rest frame, 
$g_i=2 J_i+1$ is the spin degeneracy factor, $\gamma_s \le 1$ is the 
(optional) strangeness suppression factor, $n_i^s$ is the number of strange 
quarks and antiquarks in a hadron $i$. The signs $\pm$ in the denominator 
account for the quantum statistics of a fermion or a boson, 
respectively. Then the particle number density $\rho_{i}^{\rm eq}(T, \mu_i)$
can be represented in the form of a fast converging 
series~\cite{Amelin:2006qe}, that is, 
\begin{equation}
\label{GC4} \rho_{i}^{\rm eq}(T, \mu_i) = \frac{g_i}{2 \pi^2}m^2_iT\sum_{k=1}^{\infty}
\frac{(\mp)^{k+1}}{k} \exp(\frac{k\mu_i}{T})K_2(\frac{km_i}{T})~,
\end{equation}
where $K_2$ is the modified Bessel function of the second order and $m_i$ is 
the particle mass. 

The mean multiplicity $\bar{N}_i$ of a hadron species $i$ crossing the space-like 
freeze-out hypersurface $\sigma(x)$ in Minkowski space is computed using 
effective thermal volume approximation~\cite{RKT,Nukleonika}:
\begin{equation}
\label{M8} \bar{N}_i = \rho_{i}^{\rm eq}(T,\mu_i) \int_{\sigma(x)} 
d^3\sigma_{\mu}(x)u^{\mu}(x) = \rho_{i}^{\rm eq}(T,\mu_i) V_{\rm eff}~,
\end{equation}
where the four-vector $d^3\sigma_{\mu}(x)=n_{\mu}(x)d^3\sigma(x)$ is the 
element of the freeze-out hypersurface directed along the hypersurface normal 
unit four-vector $n^{\mu}(x)$ with a positively defined zero component
($n^0(x)>0$), $d^3\sigma(x)=|d^3\sigma_{\mu}d^3\sigma^{\mu}|^{1/2}$ is
the invariant measure of this element. The probability that the produced grand 
canonical ensemble consists of $N_i$ particles is thus given by Poisson 
distribution around its mean value $\bar{N}_i$:
\begin{equation}
P(N_i) = \exp{(-\bar{N}_i)}\frac{(\bar{N}_i)^{N_i}}{N_i!}~.
\end{equation}
The chemical potential $\mu_i$ for any particle species $i$ at the chemical 
freeze-out is entirely determined by chemical potentials $\widetilde{\mu_{q}}$ per 
unit charge, i.e., per unit baryon number $B$, strangeness $S$, electric charge 
(isospin) $Q$, charm $C$, etc. It can be expressed as a scalar product, 
$\mu_i =\vec{q_i}\vec{\widetilde{\mu}}$, where 
$\vec{q_i}=\{ B_i, S_i, Q_i, C_i, ...\}$ and $\vec{\widetilde{\mu}} =
 \{\widetilde{\mu}_{B}, \widetilde{\mu}_{S}, \widetilde{\mu}_{Q}, 
 \widetilde{\mu}_{C},..\}$. Assuming constant temperature and chemical potentials 
on the chemical freeze-out hypersurface, the total quantum numbers 
$\vec{q}=\{B, S, Q,C,...\}$ of the thermal part of produced hadronic system with 
corresponding $V_{\rm eff}$ can be calculated as 
$\vec{q}=V_{\rm eff} \sum_{i}\rho_{i}^{\rm eq}\vec{q}_i$. Thus, the potentials 
$\widetilde{\mu}_{q}$ are not independent, so taking into account baryon, 
strangeness and electrical charges only and fixing the total strangeness $S$ and 
the total electric charge $Q$, $\widetilde{\mu}_{S}$ and $\widetilde{\mu}_{Q}$ can 
be expressed through baryonic potential $\widetilde{\mu}_{B}$. Therefore, the mean 
number of each particle and resonance species at chemical freeze-out is  
determined solely by the temperature $T$ and the baryonic chemical potential 
$\widetilde{\mu}_B$, which can be related within the thermal statistical approach 
using the following parameterization~\cite{Cleymans:2005xv}:
\begin{equation}
\label{Cleymans2}
T(\widetilde{\mu}_{B}) = a - b\widetilde{\mu}^2_B - c\widetilde{\mu}^4_B~,~~~~
\widetilde{\mu}_{B}(\sqrt s_{NN}) = \frac{d}{1+e\sqrt s_{NN}},
\end{equation}
where $a = 0.166 \pm 0.002$~GeV, $b = 0.139 \pm 0.016$~GeV$^{-1}$,
$c = 0.053 \pm 0.021$~GeV$^{-3}$ and
$d = 1.308 \pm 0.028$~GeV, $e = 0.273 \pm 0.008$~GeV$^{-1}$.

Since the particle densities at the chemical freeze-out
stage may be too high to consider the particles as free streaming 
(see, e.g.,~\cite{Sinyukov02}), the assumption of the common chemical and 
thermal freeze-outs can hardly be justified, so a more 
complicated scenario with different chemical and thermal freeze-outs is
implemented in HYDJET++ ($T^{\rm ch} \ge T^{\rm th}$). Within the concept of 
chemically frozen evolution, particle numbers are assumed to be conserved
except for corrections due to decay of some part of the short-lived resonances 
that can be estimated from the assumed chemical to thermal freeze-out evolution 
time. Then to determine the particle densities 
$\rho_{i}^{\rm eq}(T^{\rm th}, \mu_{i}^{\rm th})$ at the 
temperature of thermal freeze-out $T^{\rm th}$, the conservation of the 
particle fractions from the chemical to thermal freeze-out evolution time 
is supposed, and an additional input model parameter, effective pion chemical 
potential $\mu_{\pi}^{\rm eff~th}$ at thermal freeze-out, is introduced: 
\begin{equation}
 \label{GC5}\frac {\rho_{i}^{\rm eq}(T^{\rm ch}, \mu_i)}
 {\rho_{\pi}^{\rm eq}(T^{\rm ch},\mu_i^{\rm ch})}
= \frac{\rho_{i}^{\rm eq}(T^{\rm th}, \mu_i^{\rm th})}
{\rho_{\pi}^{\rm eq}(T^{\rm th}, \mu_{\pi}^{\rm eff~th})}~.
\end{equation}
Assuming for the particles heavier than pions the
Boltzmann approximation in (\ref{GC4}), one deduces from 
(\ref{GC4}) and (\ref{GC5}) the chemical potentials of hadrons at 
thermal freeze-out:
\begin{equation}
\label{GC7c} \mu_{i}^{\rm th} = T^{\rm th}\ln \left( \frac{\rho_{i}^{\rm
eq} ( T^{\rm ch}, \mu_{i}^{\rm ch})}{\rho_{i}^{\rm eq}(T^{\rm
th},\mu_i=0)}\frac{\rho_{\pi}^{\rm eq} ( T^{\rm th},
\mu_{\pi}^{\rm eff~th})}{\rho_{\pi}^{\rm eq}(T^{\rm ch},\mu_i^{\rm
ch})}\right)~.
\end{equation}
Note that it can no longer be expressed in the form 
$\mu_i =\vec{q}_i\vec{\widetilde{\mu}}$, which is valid only for chemically 
equilibrating systems.

At relativistic energies, because of the dominant longitudinal motion, it is 
convenient to parameterize the fluid flow four-velocity $\{u^0(x),\vec{u}(x)\}=
\gamma\{1,\vec{v}(x)\}$ at a point $x$ in terms of the longitudinal ($z$) and 
transverse ($r_\perp$) fluid flow rapidities
\begin{equation}
\label{HC2}\eta^u(x) = \frac{1}{2}
\ln{\frac{1+v_z(x)}{1-v_z(x)}},~~
\rho^u(x) = \frac{1}{2}
\ln{\frac{1+v_{\perp}(x)\cosh\eta^u(x)}{1-v_{\perp}(x)\cosh\eta^u(x)}}~,
\end{equation}
where $v_{\perp} = | \vec{v}_{\perp}|$ is the magnitude of the transverse component of 
the flow three-velocity $ \vec{v}=\{\vec{v}_{\perp},v_z \} = 
\{ v_\perp \cos \phi^u , v_\perp \sin \phi^u, v_z\}$, i.e.,
\begin{equation}
\begin{array}{c}
\label{HC6}u^{\mu}(x)=
\{\cosh \rho^u \cosh \eta^u, \sinh \rho^u \cos \phi^u, \sinh \rho^u \sin \phi^u,
\cosh \rho^u \sinh \eta^u \}
\\
=\{
 (1+u_{\perp}^{2})^{1/2} \cosh \eta^u,\vec{u}_{\perp}, (1+u_{\perp}^{2})^{1/2} \sinh \eta^u\}~,
\end{array}
\end{equation}
$\vec{u}_{\perp} = \gamma\vec{v}_{\perp}= \cosh\eta^u\gamma_{\perp}
\vec{v}_{\perp}$, $\gamma_{\perp}=\cosh \rho^u$. The linear transverse rapidity 
profile is used and the momentum anisotropy parameter $\delta (b)$ is 
introduced here: 
\begin{eqnarray}
\label{URBaN}
& & u^{x} = \sqrt{1+\delta(b)} \sinh \tilde \rho_u\cos{\phi}~,~~
u^{y} = \sqrt{1-\delta(b)} \sinh \tilde \rho_u\sin{\phi}~,\nonumber \\
& & \tilde \rho_u=\frac{r}{R_f(b)}\rho_u^{\rm max}(b=0)~,
\end{eqnarray}
where $\phi$ is the spatial azimuthal angle of the fluid element, $r$ is its
radial coordinate, $\rho_u^{\rm max}(b=0)$ is the maximal transverse flow 
rapidity for central collisions, and $R_f (b)$ is the mean-square radius of the 
hadron emission region~\cite{Amelin:2007ic}, which determines the fireball 
transverse radius $R(b,\phi)$ in the given azimuthal direction $\phi$ through 
the spatial anisotropy parameter $\epsilon(b)$:
\begin{equation}
\label{Rbphi}
R(b,\phi)= R_f(b)
 \frac{\sqrt{1-\epsilon^2(b)}} {\sqrt{1+\epsilon(b) \cos2\phi}}~.
\end{equation}
For this case, it is straightforward to derive the formula for the total 
effective volume of hadron emission from the hypersurface of proper time 
$\tau$=const~\cite{Amelin:2006qe,Amelin:2007ic}: 
\begin{eqnarray}
\label{HD3} V_{\rm eff}=\int\limits_{\sigma(t, \vec{x})} d^3\sigma_{\mu}(t, 
\vec{x})u^{\mu}(t, \vec{x}) = \tau \int\limits_0^{2\pi} d\phi \int\limits_0^{R(b,\phi)} 
(n_{\mu}u^{\mu}) r dr \int\limits_{\eta_{\min}}^{\eta_{\max}} f(\eta) d\eta,
\end{eqnarray}
where $(n_{\mu}u^{\mu})=\cosh \tilde \rho_u
\sqrt{1+\delta(b) \tanh^2 \tilde \rho_u \cos 2 \phi}$, and $f(\eta)$ is the 
longitudinal flow rapidity profile (Gaussian or uniform). 

The value $V_{\rm eff}$ is calculated in HYDJET++ only for central collisions
($b=0$), and then for non-central collisions it is supposed to be 
proportional to the mean number of nucleons-participants 
$\overline{N_{\rm part}}(b)$ (\ref{npart}): 
\begin{equation}
\label{veff-b}
V_{\rm eff}(b)=V_{\rm eff}(b=0)\frac{\overline{N_{\rm part}}(b)}
{\overline{N_{\rm part}}(b=0)}~.
\end{equation}
Since the fireball transverse radius $R_f(b=0)$ and the freeze-out proper time
$\tau _f(b=0)$ (as well as its standard deviation $\Delta \tau _f(b=0)$ -- the 
emission duration) are the input model parameters, it is straightforward to 
determine these values for non-central collisions through the effective volume 
$V_{\rm eff}$ and input anisotropy parameters $\epsilon$ and $\delta$ at the given 
$b$: 
\begin{eqnarray}
\label{r-b}
& & R_f(b)=R_f(0)\left( \frac{V_{\rm eff}[R_f(0),\tau_f(0),\epsilon (0), \delta (0)]}
{V_{\rm eff}[R_f(0),\tau_f(0),\epsilon (b), \delta (b)]}\right) ^{1/2}
\left( \frac{\overline{N_{\rm part}}(b)}{\overline{N_{\rm part}}(b=0)}
\right)^{1/3},\\
\label{tau-b}
& & \tau _f(b)= \tau _f(0) \left( \frac{\overline{N_{\rm part}}(b)}
{\overline{N_{\rm part}}(b=0)} \right)^{1/3},~\Delta \tau _f(b)= \Delta \tau _f(0) 
\left( \frac{\overline{N_{\rm part}}(b)} {\overline{N_{\rm part}}(b=0)}
\right)^{1/3}.
\end{eqnarray} 
Note that such choice of centrality dependence for  $R_f(b)$ and $\tau_f(b)$ is 
inspired by the experimentally observed dependence of momentum correlation 
radii $R_{\rm side}(b)$, $R_{\rm out}(b)$, $R_{\rm long}(b)$ $\propto 
N_{\rm part}(b)^{1/3}$~\cite{Adams:2003ra,Adler:2004rq}. The detailed study of 
femtoscopic momentum correlations at RHIC and LHC (including their centrality 
dependence) is planned in the HYDJET++ frameworks for the future.  

The momentum and spatial anisotropy parameters $\delta (b)$ and $\epsilon (b)$ 
can be treated independently for each centrality, or can be related to each 
other through the dependence of the elliptic flow coefficient 
$v_2(\epsilon, \delta)$ (the second-order Fourier coefficient in the hadron 
distribution over the azimuthal angle $\varphi$ relatively to the reaction 
plane) obtained in the hydrodynamical approach~\cite{Wiedemann98}:
\begin{equation}
\label{v2-eps-delta1}
v_2 \propto \frac{2(\delta-\epsilon)}{(1-\delta^2)(1-\epsilon^{2})}~. 
\end{equation}
Then using the predicted by hydrodynamical models (and observed at RHIC) 
proportionality of the value $v_2(b)$ and the 
initial ellipticity $\epsilon_0 (b)=b/2R_A$ 
(i.e. $v_2 \propto \epsilon_0$), one can derrive the
relation between $\delta(b)$ and $\epsilon (b)$ (supposing also
$\epsilon \propto \epsilon_0$):
\begin{equation}
\label{v2-eps-delta2}
\delta = \frac{\sqrt{1+4B(\epsilon+B)}-1}{2B}~,~~~B=C(1-\epsilon^2)\epsilon~,
~~~\epsilon=k \epsilon_0,
\end{equation}
where $C$ and $k$ are the independent on centrality coefficients, which should be specified instead of $\delta (b)$ and $\epsilon (b)$ dependences. The best fit of RHIC data to the elliptic flow (Figure \ref{fig_v2_all} in Section 6) is obtained with the values $C=2$ and $k=0.175$.  

The ``thermal'' hadronic state in HYDJET++ consists of stable hadrons and 
resonances produced from the SHARE particle data table~\cite{share}, which contains 
360 particles (excluding any not well established resonance states). The decays of 
resonances are controlled by the decay lifetime 
$1/\Gamma$, there $\Gamma$ is the resonance width specified in the particle table, 
and these decays occur with the probability density $\Gamma \exp(-\Gamma \tau)$ in 
the resonance rest frame. Then the decay products are boosted to the reference 
frame in which the freeze-out hypersurface was defined. The space-time 
coordinates of the decaying particle are shifted from its initial position on 
the decay length $\Delta \tau P/M $ ($M$ and $P$ are the decaying particle 
mass and four-momenta respectively). The branching ratios are also taken from 
the SHARE~\cite{share}. Only the two- and three-body decays are taken into 
account in the model. The cascade decays are also possible. Resonances are 
given the mass distribution according to a non-relativistic Breit-Wigner 
\begin{equation}
\label{nrBW}
P(m) dm \propto \frac{1}{(m-m_0)^2+\Delta m^2/4}dm,
\end{equation}
where $m_0$ and $\Delta m$ are the resonance nominal mass and width 
respectively. The Breit-Wigner shape is truncated symmetrically, 
$\mid m-m_0\mid<\Delta m$, with $\Delta m$ taken for each particle from the
PYTHIA~\cite{pythia} ($\Delta m=0$ for some narrow resonances which are not 
present in PYTHIA). 

\section{Simulation procedure} 

Before any event generation, HYDJET++ run starts from PYTHIA initialization at the 
given c.m.s. energy per nucleon pair $\sqrt{s}$ (input parameter \verb*|fSqrtS|), 
and then it calculates the total inelastic NN cross section 
$\sigma _{NN}^{\rm in}(\sqrt{s})$ (output parameter \verb*|Sigin|) and the hard 
scattering NN cross section $\sigma_{NN}^{\rm hard}(\sqrt{s},p_T^{\rm min})$ 
(output parameter \verb*|Sigjet|) with the minimum transverse momentum transfer 
$p_T^{\rm min}$ (input parameter \verb*|fPtmin|). Then the tabulation of nuclear 
thickness function $T_A(b)$ and nuclear overlap function $T_{AA}(b)$ is performed.  
If the impact parameter $b$ of heavy ion AA collision is not fixed (input parameter 
\verb*|fIfb| $\ne 0$), its value $b$ (output parameter \verb*|Bgen|) is generated in 
each event between the minimum (input parameter \verb*|fBmin|) and maximum (input 
parameter \verb*|fBmax|) values in accordance with the differential inelastic AA 
cross section, obtained from the standard generalization of Glauber multiple 
scattering model~\cite{glauber} to the case of independent inelastic 
nucleon-nucleon collisions:   
\begin{equation}
\label{sigin_b} 
\frac {d^2 \sigma^{AA}_{\rm 
in}}{d^2b} (b, \sqrt{s}) = \left[ 1 - \left( 1- \frac{1}  
{A^2}T_{AA}(b) \sigma^{\rm in}_{NN} (\sqrt{s}) \right) ^{A^2} \right]~.   
\end{equation} 
If the impact parameter $b$ is fixed (\verb*|fIfb| $=0$), then its value 
\verb*|Bgen| in each event is equal to the input parameter \verb*|fBfix|. After 
specification of $b$ for each given event, the mean numbers of binary NN sub-collisions 
$\overline{N_{\rm bin}}$ (output parameter \verb*|Nbcol|) and nucleons-participants 
$\overline{N_{\rm part}}$ (output parameter \verb*|Npart|) are calculated:   
\begin{eqnarray} 
\label{nbcol}
& & \overline{N_{\rm bin}}(b,\sqrt{s}) = T_{AA}(b) \sigma _{NN}^{\rm in}(\sqrt{s})~, 
\\  
\label{npart}
& & \overline{N_{\rm part}}(b,\sqrt{s}) = \int\limits_0^{2\pi} d\psi 
\int\limits_0^{\infty} rdrT_A(r_1) 
\left[ 1-\exp{\{ \sigma^{\rm in}_{NN}(\sqrt{s}) T_A(r_2)\} } \right]~.  
\end{eqnarray} 
The next step is the simulation of particle production proper in the event. 
The soft, hydro-type state and the hard, multi-jet state are simulated 
independently. When the generation of soft and hard states in each event at given 
$b$ is completed, the event record (information about coordinates and momenta of 
initial particles, decay products of unstable and stable particles) is 
formed as the junction of these two independent event outputs in the 
\verb*|RunOutput.root| file. 

\subsection{Generation of the hard multi-jet state}

The following event-by-event MC simulation procedure is applied to generate the 
hard state resulting from multi-parton fragmentation in the case when jet 
production is switched on (input parameter \verb*|fNhsel|=1, 2, 3 or 4). 

\begin{enumerate}

\item Calculation of the number of NN sub-collisions $N_{AA}^{\rm jet}$ (output 
parameter \verb*|Njet|) producing hard parton-parton scatterings of selected type 
(QCD-dijets by default, PYTHIA parameter \verb*|msel=1|) with $p_T>p_T^{\rm min}$, 
according to the binomial distribution around its mean value (\ref{numjets}) (without 
shadowing correction yet, $S=1$). For this purpose, each of \verb*|Nbcol| 
sub-collisions is tested basing on comparison of random number $\xi_i$ generated 
uniformly in the interval $[0,1]$ with the probability \verb*|pjet=Sigjet/Sigin| to 
produce the hard process. The $i$-th sub-collision is accepted if $\xi_i<$\verb*|pjet|, 
and is rejected in the opposite case. 

\medskip

\item Selecting the type of hard NN sub-collision (pp, np or nn) in accordance with 
the phenomenological formula for the number of protons $Z$ in the stable nucleus A, 
$Z=A/(1.98+0.015A^{2/3})$.  For this purpose, each of \verb*|Njet| ``successful'' 
sub-collisions is tested basing on comparison of two random numbers $\xi^1_i$ and 
$\xi^2_i$ generated uniformly in the interval $[0,1]$ with the probability $Z/A$. 
The proton-proton sub-collision is selected if $\xi^1_i,\xi^2_i<Z/A$,
neutron-neutron sub-collision --- if $\xi^1_i,\xi^2_i>Z/A$, and proton-neutron  
sub-collision --- in other cases.  

\medskip

\item Generation of multi-parton production in \verb*|Njet| hard NN 
sub-collisions by calling PYTHIA \verb*|Njet| times (\verb*|call pyinit| and 
\verb*|call pyevnt|, parton fragmentation being switched off by setting 
\verb*|mstp(111)=0|). The spatial vertex of jet production for each sub-collision 
is generated by PYQUEN routine according to the distribution
\begin{equation} 
\label{vertex}
\frac{dN^{\rm jet}}{d\psi r dr} (b) = \frac{T_A(r_1)\cdot T_A(r_2)}{T_{AA}(b)}~. 
\end{equation} 

\medskip

\item If jet quenching is switched on (\verb*|fNhsel|=2 or 4), initial 
PYTHIA-produced partonic state is modified by QCD-medium effects with PYQUEN 
generator for each sub-collision separately. The radiative and collisional energy 
loss are both taken into account by default (input parameter \verb*|fIenglu|=0), but 
the options to have only radiative loss (\verb*|fIenglu|=1) or only collisional loss 
(\verb*|fIenglu|=2) are envisaged. For each hard parton with the initial transverse 
momentum $p_T>3$ GeV/$c$ and pseudorapidity $\mid\eta\mid <3.5$ in the given
sub-collision, the following rescattering scheme is applied. 

\smallskip

\noindent
$\bullet$ Calculation of the scattering cross section $\sigma (\tau_i) = 
\int dt~d\sigma/dt$ and generation of the transverse momentum transfer 
$t (\tau_i)$ in the $i$-th scattering according to (\ref{sigt}) ($\tau_i$ is a 
current proper time). \\
$\bullet$ Generation of the displacement between the $i$-th and $(i+1)$-th
scatterings, $l_i = (\tau_{i+1} - \tau_i)$:  
\begin{equation} 
\frac{dP}{dl_i} = \lambda^{-1}(\tau_{i+1}) \exp{(-\int\limits_0^{l_i}
\lambda^{-1} (\tau_i + s)ds)} ~,~~ \lambda^{-1}(\tau ) =\sigma (\tau ) \rho 
(\tau )~,
\end{equation}  
and calculation of the corresponding transverse distance, $l_i p_T/E$. \\
$\bullet$ Generation of the energy of a radiated in the $i$-th scattering 
gluon, $\omega _i=\Delta E_{{\rm rad},i}$, according to (\ref{radiat}) and 
(\ref{radmass}):
\begin{eqnarray}
& &\frac{dI}{d\omega }| _{m_q=0} = \frac{2 \alpha_s (\mu_D^2) 
\lambda C_R}{\pi L \omega } \left[ 1 - y + \frac{y^2}{2} \right] 
\>\ln{\left| \cos{(\omega_1\tau_1)} \right|}~,\\ 
& & \frac{dI}{d\omega }| _{m_q \ne 0} =  \frac{1}{(1+(\beta \omega )^{3/2})^2}
\frac{dI}{d\omega }| _{m_q=0}~,   
\end{eqnarray}
and its emission angle $\theta _i$ relative to the parent parton determined  
according to the small-angle distribution (\ref{sar}) (input parameter 
\verb*|fIanglu|=0). The options to have wide-angle (\verb*|fIanglu|=1) or 
collinear (\verb*|fIanglu|=2) distributions are also envisaged. \\
$\bullet$ Calculation of the collisional energy loss in the $i$-th scattering 
\begin{equation} 
\Delta E_{{\rm col},i} = \frac{t_i}{2 m_0}~,
\end{equation} 
where energy of ``thermal'' medium parton $m_0$ is generated according to the 
isotropic Boltzmann distribution at the temperature $T(\tau _i)$. \\
$\bullet$ Reducing the parton energy by collisional and radiative loss per 
each $i$-th scattering,
\begin{equation} 
\Delta E_{{\rm tot},i} = \Delta E_{{\rm col},i} + \Delta E_{{\rm rad},i}~, 
\end{equation}
and changing the parton momentum direction by means of adding the transverse momentum 
recoil due to elastic scattering $i$,
\begin{equation} 
\Delta k_{t,i}^2 =(E-\frac{t_i}{2m_{0i}})^2-(p-\frac{E}{p}\frac{t_i}{2m_{0i}}-
\frac{t_i}{2p})^2-m_p^2~.
\end{equation} 
$\bullet$ Going to the next rescattering, or halting the rescattering if one of the
following two conditions is fulfilled: {\em (a)} the parton escapes the hot QGP 
zone, i.e. the temperature in the next point $T(\tau_{i+1},r_{i+1},\eta_{i+1})$ 
becomes lower than $T_c=200$ MeV; or {\em (b)} the parton loses so much of energy 
that its transverse momentum $p_T (\tau_{i+1})$ drops below the average transverse 
momentum of the ``thermal'' constituents of the medium, $2T (\tau_{i+1})$.  In the 
latter case, such a parton is considered to be ``thermalized" and its momentum in 
the rest frame of the quark-gluon fluid is generated from the random ``thermal'' 
distribution, $dN/d^3p \propto \exp{\left( -E/T\right) }$, boosted to the 
center-of-mass of the nucleus-nucleus collision.

\smallskip

\noindent
At the end of each NN sub-collision, adding new (in-medium emitted) gluons to 
the PYTHIA parton list and rearrangement of partons to update string formation 
with the subroutine PYJOIN are performed. An additional gluon is included in the 
same string as its ``parent'', and colour connections of such gluons are
re-ordered relative to their $z$-coordinates along the string. 

\medskip

\item If the nuclear shadowing is switched on (input parameter \verb*|fIshad|=1) and 
beam ions are Pb, Au, Pd or Ca (\verb*|fAw|$=$\verb*|207.|, \verb*|197.|, \verb*|110.| 
or \verb*|40.|), each hard NN sub-collision is tested basing on comparison of random 
number $\xi_i$ generated uniformly in the interval $[0,1]$ with the shadowing 
factor $S$ (\ref{S_shad}) taken from the available parameterization 
(\verb*|subroutine ggshad|). It is determined by the type of initially 
scattered hard partons, momentum fractions obtained by the partons at the initial 
hard interaction $x_{1,2}$ (PYTHIA parameters \verb*|pari(33)|, 
\verb*|pari(34)|), the square of transverse momentum transfer in the hard 
scattering $Q^2$ (PYTHIA parameter \verb*|pari(22)|), and the transverse 
position of jet production vertex relative to the centres of nuclei $r_{1,2}$. 
The given sub-collision is accepted if $\xi_i<S$, and is rejected in the 
opposite case. 

\medskip

\item Formation of hadrons for each ``accepted'' hard NN 
sub-collision with PYTHIA (parton fragmentation being switched on by 
\verb*|call pyexec|), and final junction of \verb*|Njet| sub-events to common 
array \verb*|hyjets| using standard PYTHIA event output format. If particle
decays in PYTHIA are switched off (input PYTHIA parameter \verb*|mstj(21)=0|), then 
formation of final hadrons from the two- and three-body decays of resonances follows 
the same pattern as that for soft hydro-part with the branching ratios 
taken from the SHARE particle decay table~\cite{share}. 

\end{enumerate}

\subsection{Generation of the soft ``thermal'' state}

The following event-by-event MC simulation procedure is applied to 
generate the soft ``thermal'' state in the case when soft hadroproduction is switched 
on (input parameter \verb*|fNhsel|=0, 1, or 2). 

\begin{enumerate}

\item Initialization of the chemical freeze-out parameters. It includes the 
calculation of particle number densities according to (\ref{GC4}) 
(the calculation of chemical freeze-out temperature, baryon potential and 
strangeness potential as a function of beam c.m.s. energy \verb*|fSqrtS| 
according to (\ref{Cleymans2}) is performed in advance if the corresponding flag 
\verb*|fTMuType|$>$0). So far, only the stable hadrons and resonances consisting 
of $u$, $d$, $s$ quarks are taken into account from the SHARE particle data 
table~\cite{share}.  

\medskip

\item Initialization of the thermal freeze-out parameters (if $T^{\rm th}<
T^{\rm ch}$). It includes the calculation of chemical potentials according 
to (\ref{GC7c}) and particle number densities according to (\ref{GC5}). 

\medskip

\item Calculation of the effective volume of hadron emission region 
$V_{\rm eff}(b)$  according to (\ref{veff-b}) and (\ref{HD3}), the fireball 
transverse radius $R_f(b)$, the freeze-out proper time $\tau _f(b)$ and the 
emission duration $\Delta \tau _f(b)$  according to 
(\ref{r-b}) and (\ref{tau-b}), and the mean multiplicity of each particle
species according to (\ref{M8}). Then the multiplicity is generated around its 
mean value according to the Poisson distribution (\ref{GC5}).

\medskip

\item For each hadron the following procedure to generate its four-momentum is 
applied (resonance mass is generated according to (\ref{nrBW})). 

\smallskip

\noindent
$\bullet$  Generation of four-coordinates of a hadron in the fireball rest frame 
$x^{\mu}=\{\tau \cosh{\eta},~r \cos{\phi},~r \sin{\phi},~\tau \sinh{\eta}\}$ on
each freeze-out hypersurface segment $\tau(r)$ for the element
$d^3\sigma_{\mu}u^{\mu}=d^3\sigma_0^*=n_0^*(r) \mid 1-(d\tau/dr)^2 \mid ^{1/2}\tau
(r)d^2rd\eta$, assuming $n_0^*$ and $\tau$ to be the functions of r (i.e., independent  of $\eta$, $\phi$). It includes sampling uniformly distributed $\phi$ in the interval 
$[0, 2\pi]$, generating $\eta$ according to the uniform distribution in the 
interval $[-\eta_{\rm max},\eta_{\rm max}]$ (if flag \verb*|fEtaType|$\le$0) or
Gaussian distribution $\exp(-\eta^2/2\eta_{\rm max}^2)$ (if flag
\verb*|fEtaType|$>$0) and $r$ in the interval $[0, R_f(b)]$) using a $100$\% 
efficient procedure similar to the ROOT routine \verb*|GetRandom()|.\\
$\bullet$ Calculation of the corresponding collective flow four-velocities 
according to (\ref{HC6}) and (\ref{URBaN}).\\
$\bullet$ Generation of the three-momentum of a hadron in the fluid element 
rest frame $p^*\{\sin \theta_p^*cos \phi_p^*,~\sin \theta_p^*sin 
\phi_p^*,~\cos \theta_p^*\}$ according to the equilibrium distribution function 
$f_i^{\rm eq}(p^{0*};T,\mu_i)p^{*2}dp^{*}d\cos\theta_p^* d\phi_p^*$ by means of sampling uniformly distributed $\cos\theta_p^*$ in the interval $[-1,1]$ and 
$\phi_p^*$ in  the interval $[0, 2\pi]$, and generating $p^*$ using a $100$\% 
efficient procedure (similar to ROOT routine \verb*|GetRandom()|).\\
$\bullet$ The standard von Neumann rejection/acceptance procedure to take into 
account the difference between the true probability 
$W_{\sigma,i}^* d^3\sigma d^3\vec{p}^{*}/p^{0*}$ and the probability 
$n^{0*} f_i^{\rm eq}(p^{0*};T,\mu_i)d^2\vec{r} d\eta d^3\vec{p}^{*}$
corresponding to the previous simulation steps. 
For this purpose, the residual weight is calculated~\cite{Amelin:2006qe}:  
\begin{equation}
\label{wres2}
W_i^{\rm res}=\frac{W_{\sigma,i}^*d^3\sigma}{n^{0*}p^{0*}f_i^{\rm eq}d^2\vec{r} 
d\eta}=\tau\left(1-\frac{\vec{n^*}\vec{p^*}}{n^{0*}p^{0*}}\right)~. 
\end{equation}
Then the simulated hadron four-coordinate and four-momentum is tested basing on 
comparison of $W_i^{\rm res}$ with the random number $\xi_i$ generated uniformly 
in the interval $[0,\max(W_i^{\rm res})]$. The $i$-th hadron is accepted if 
$\xi_i<W_i^{\rm res}$, and rejected in the opposite case (then the generation of
its four-coordinate and four-momentum is repeated).\\ 
$\bullet$ Boost of the hadron four-momentum in the center mass frame of the event 
using the velocity field $\vec{v}(x)$, that is, 
\begin{equation}
\label{M69} 
p^{0} = \gamma(p^{0*} + \vec{v}\vec{p}^{~*})~,~~~~~
\vec{p} = \vec{p}^{~*} + \gamma (1 + \gamma)^{-1} (p^{0*} +p^0 )\vec{v}~.
\end{equation} 
Note that the high generation speed for this algorithm is achieved due to  
almost 100\% generation efficiency because of nearly uniform  
residual weights $W_i^{\rm res}$ (\ref{wres2}).

\medskip 

\item Formation of final hadrons from the two- and three-body decays of 
resonances with the random choice of decay channel according to the branching 
ratios taken from the particle data file \verb*|tabledecay.txt| (by default, if 
flags \verb*|fDecay| and \verb*|fWeakDecay| $\ge$0). 
The particle ``decay'' programs developed by N.S.~Amelin for FAST MC generator~\cite{Amelin:2006qe,Amelin:2007ic} are implemented in HYDJET++. \\ 
$\bullet$ For the two-body decay, the momenta $|p_{1,2}|$ of decay products 
in the resonance rest frame is calculated as: 
\begin{equation}
\label{d2}
\mid p_{1,2}\mid  = 0.5 \sqrt{((M^{2} - m_{1}^{2} - m_{2}^{2})^{2} - 4  m_{1}^{2}  
m_{2}^{2})} / M~,
\end{equation}
where $M$ is the resonance mass, and $m_{1,2}$ are the decay product  
masses. The space orientation of the $|p_{1,2}|$ is generated randomly by 
sampling uniformly distributed cosine of the polar angle $\cos{(\theta)}$ in 
the interval $[-1,1]$ and the azimuthal angle $\phi$ in the interval $[0, 2\pi]$ 
(in the resonance rest frame). The opposite directions for $p_{1,2}$ vectors are 
chosen according to momentum conservation law. \\
$\bullet$ For the three-body decay, the resonance energy is divided between the 
kinetic energy of three decay products uniformly (assuming constant matrix element of the decay):
\begin{eqnarray}
\label{d3-2}
& &  E_1^{\rm kin} = \xi_1 \Delta M~, ~~~~~~~~~~~
\mid p_1\mid = \sqrt{((E_1^{\rm kin})^{2} + 2 E_1^{\rm kin} m_1)}  \nonumber \\ 
& &  E_2^{\rm kin} = (1-\xi_2) \Delta M \nonumber~, 
~~~\mid p_2\mid = \sqrt{((E_2^{\rm kin})^{2} + 2 E_2^{\rm kin} m_2)}   \\   
& &  E_3^{\rm kin} = (\xi_2-\xi_1) \Delta M \nonumber~, ~~\mid p_3\mid = 
\sqrt{((E_3^{\rm kin})^{2} + 2 E_3^{\rm kin} m_3)}~,  
\end{eqnarray}									      
where $\xi_1$ and $\xi_2$ are the two random numbers distributed uniformly in the 
interval $[0,1]$ under the condition  $\xi_2 > \xi_1$, and 
$\Delta M =  M - m_1 - m_2 - m_3$. 
For the first particle, the space orientation of the $|p_{1}|$ is generated 
randomly by sampling uniformly distributed $\cos(\theta)$ in the interval 
$[-1,1]$ and $\phi$ in the interval $[0, 2\pi]$ (in the resonance rest frame). Then
the three-momenta of remaining two particles are calculated taking into account 
complanarity of the decay and the conservation laws:
\begin{eqnarray}
& & p_2^X= \mid p_2\mid (\sin{\theta_N}\cos{\phi_N}\cos{\theta}\cos{\phi}
 - \sin{\theta_N}\sin{\phi_N}\sin{\phi} + \nonumber \\
& & ~~~~~~~~~~~~~~\cos{\theta_N}\sin{\theta}\cos{\phi})~, \nonumber \\  
& & p_2^Y= \mid p_2\mid (\sin{\theta_N}\cos{\phi_N}\cos{\theta}\sin{\phi}
 - \sin{\theta_N}\sin{\phi_N}\cos{\phi} + \nonumber \\
& & ~~~~~~~~~~~~~~\cos{\theta_N}\sin{\theta}\sin{\phi})~, \nonumber  \\
& & p_2^Z= \mid p_2\mid (-\sin{\theta_N}\cos{\phi_N}\sin{\theta}+
 \cos{\theta_N}\cos{\theta})~, \nonumber \\
& & \vec{p_3}=-(\vec{p_2}+\vec{p_1})~, \nonumber \\
& & \cos{\theta_N} = (\mid p_2\mid ^2 - \mid p_3\mid ^2 - \mid p_1\mid ^{2})/(2 
\mid p_1\mid \mid p_3\mid )~, \nonumber 
\end{eqnarray}	
and $\phi_N$ is generated uniformly in the interval $[0, 2\pi]$. Finally, the 
momenta of decay particles for two- and three-body 
decays are boosted to the center-of-mass of the nucleus-nucleus collision with 
the resonance velocity. The space-time coordinates of the decay products coincide 
with the coordinates of decay of the initial resonance. 

\end{enumerate}

\section{Overview of HYDJET++ software structure}

\begin{figure}
\begin{center}
\includegraphics[angle=270,width=12cm]{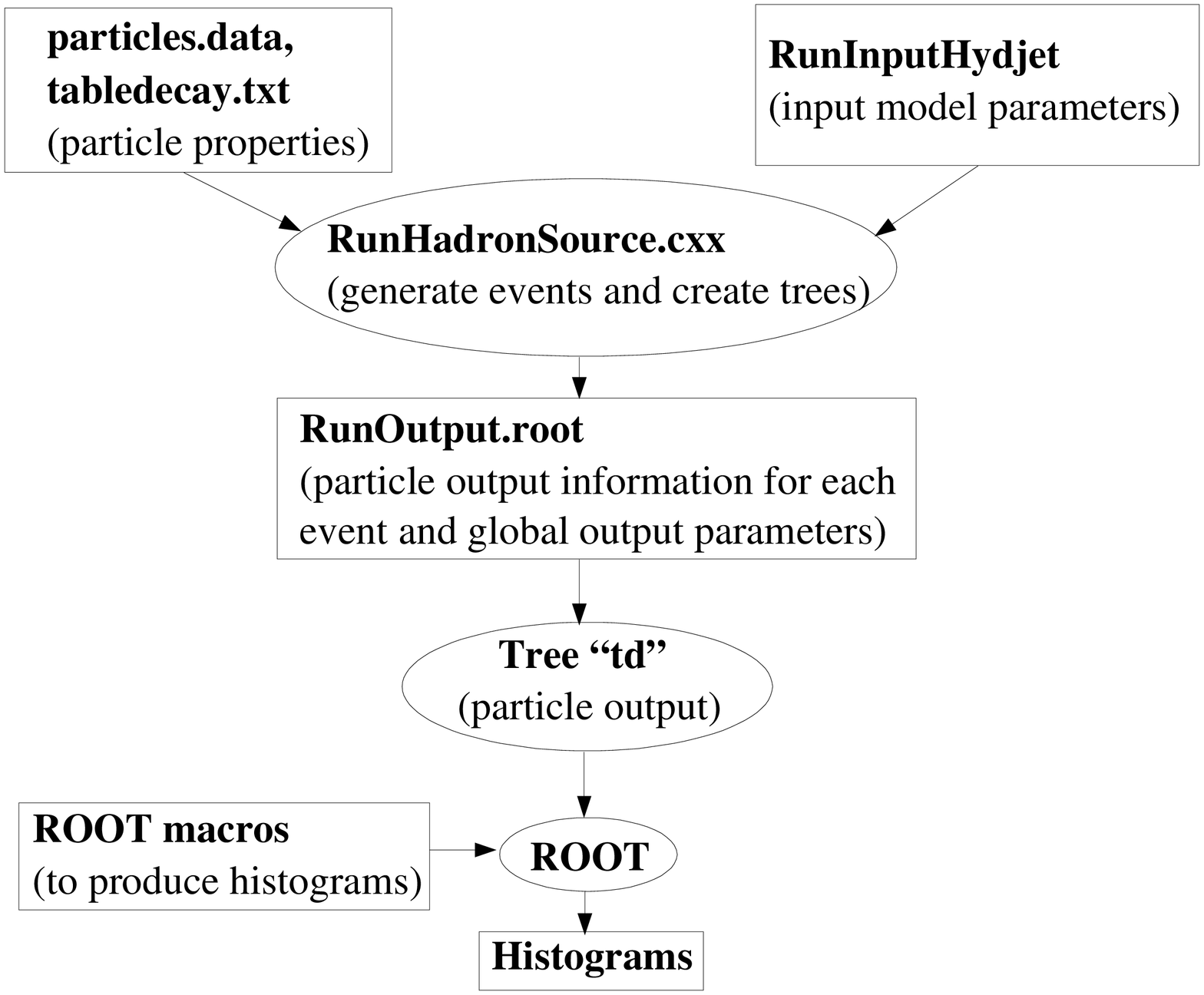}
\end{center}
\caption{The block structure of {\bf \small Hydjet++}.}
\label{fig_hydjet_block}
\end{figure}

The basic frameworks of HYDJET++ are preset by the object-oriented C++ language and 
the ROOT environment~\cite{root}. There is also the Fortran-written part~\cite{hydjet}  which is included in the generator structure as a separate directory. 
The block structure of HYDJET++ is shown 
in Figure \ref{fig_hydjet_block}. The main program elements (particle 
data files, input and output files, C++ and Fortran routines) are described in 
details in this section below.  

\subsection{Particle data files}

The information regarding the particle species included in the soft part of the 
HYDJET++ event is stored in the files \verb*|particles.data| and \verb*|tabledecay.txt|. 
The \verb*|particles.data| file contains the definition (PDG code) and physical 
properties (mass, decay width, spin, isospin, valence quark composition) of 360 
stable hadrons and resonances. The \verb*|tabledecay.txt| file contains decay channels 
and branching ratios. The structure of these files is the same as that in SHARE particle data table~\cite{share} and in event generator THERMINATOR~\cite{therminator}, where the corresponding description in more details can be found. 

The file \verb*|particles.data| has the following format. 

\noindent
\verb*|name| -- the particle label; \\
\verb*|mass| -- mass in GeV/$c^2$; \\
\verb*|width| -- width in GeV/$c^2$; \\
\verb*|spin| -- spin; \\
\verb*|I| -- isospin; \\
\verb*|I3| -- third component of isospin; \\
\verb*|q| -- number of light valence quarks in the particle; \\
\verb*|s| -- number of strange valence quarks in the particle; \\
\verb*|aq| -- number of light valence antiquarks in the particle; \\
\verb*|as| -- number of strange valence antiquarks in the particle; \\
\verb*|c| -- number of charm valence quarks in the particle; \\
\verb*|ac| -- number of charm valence antiquarks in the particle; \\
\verb*|MC| -- the particle identification code. 

The file \verb*|tabledecay.txt| has the following format. 

\noindent 
\verb*|PdgParent| -- parent particle code; \\
\verb*|PdgDaughter1| -- first particle (decay product) code; \\
\verb*|PdgDaughter2| -- second particle (decay product) code; \\
\verb*|PdgDaughter3| -- third particle (decay product) code (appears for the
three-body decays only); \\
\verb*|BR| -- the branching ratio for the decay. 

Note that the default PYTHIA$\_$6.4 particle data settings~\cite{pythia} are used 
to generate hard part of HYDJET++ event.  

\subsection{Input parameters and files}

\begin{figure}
\begin{center}
\includegraphics[angle=270,width=14.5cm]{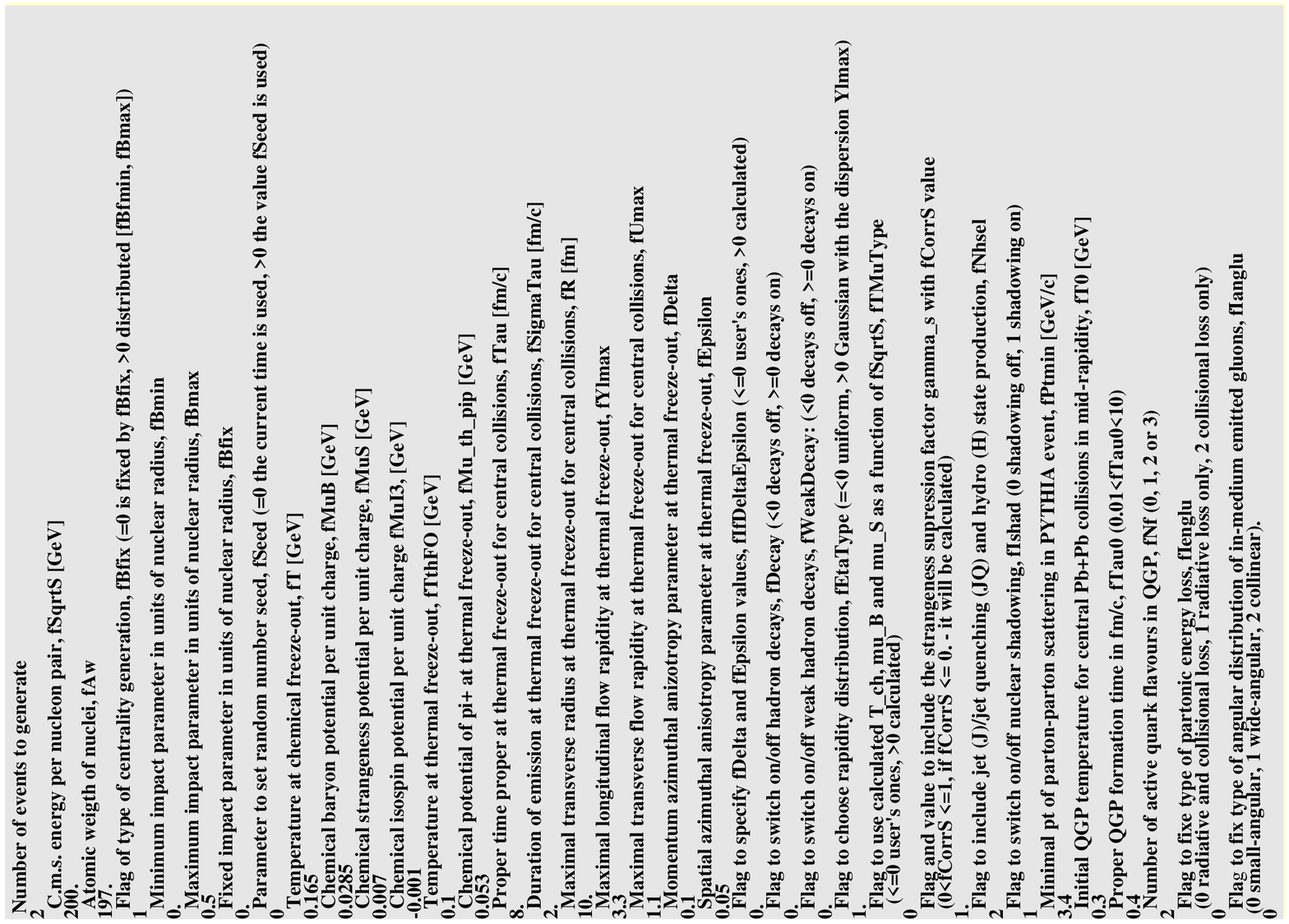}
\end{center}
\caption{The input parameter file {\bf \small RunInputHydjetRHIC200} 
(default).}
\label{fig_inrhic}
\end{figure}

\begin{figure}
\begin{center}
\includegraphics[angle=270,width=14.5cm]{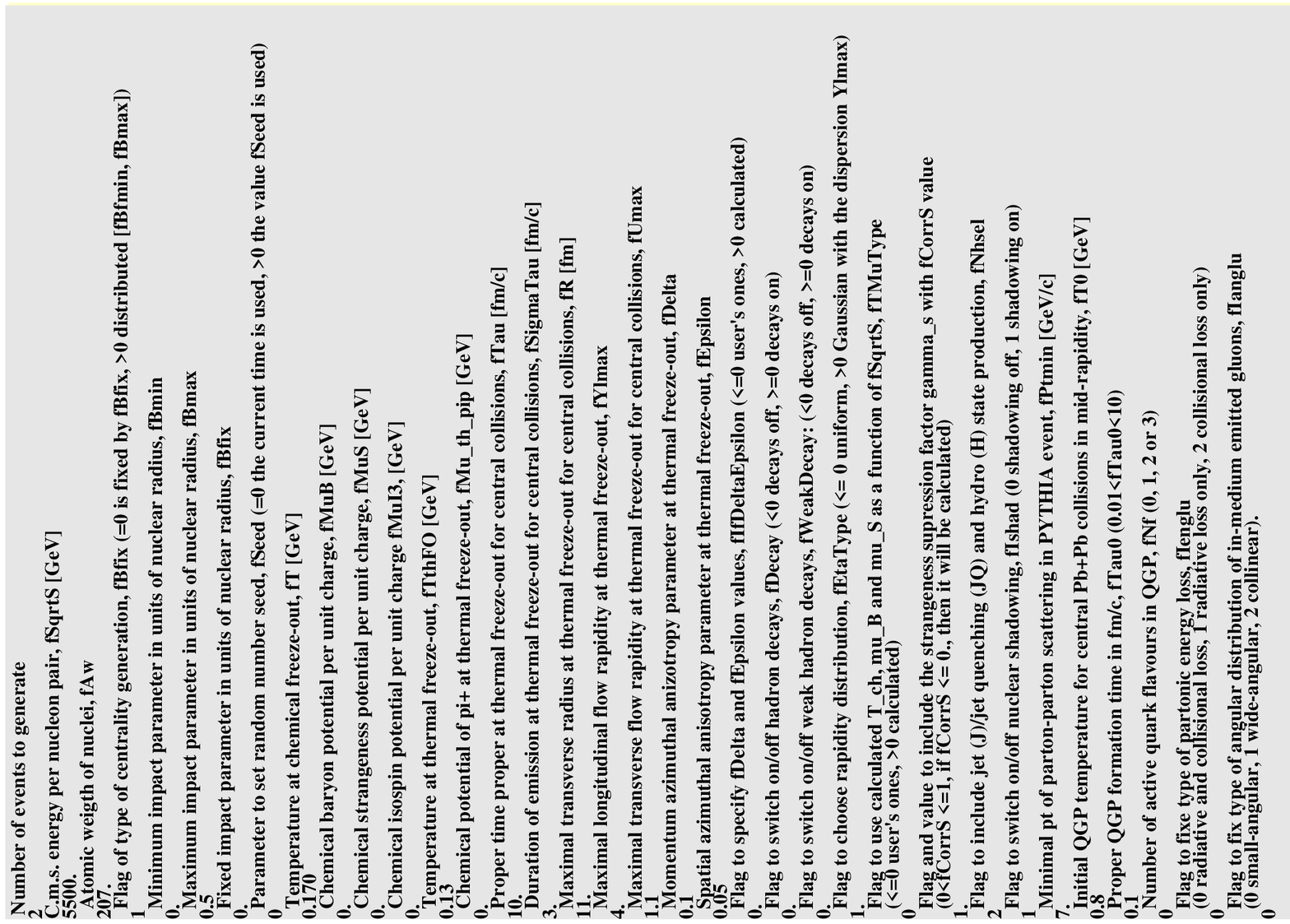}
\end{center}
\caption{The input parameter file {\bf \small RunInputHydjetLHC5500} 
(default).}
\label{fig_lhc}
\end{figure}

Run of HYDJET++ is controlled by the file \verb*|RunInputHydjet| for different 
type of input parameters. For current version of the generator, two additional 
files with the optimized parameters for Au+Au collisions at $\sqrt{s}=200 A$ 
GeV (\verb*|RunInputHydjetRHIC200|, Fig.~\ref{fig_inrhic}) and for Pb+Pb 
collisions at $\sqrt{s}=5500 A$ GeV (\verb*|RunInputHydjetLHC5500|, 
Fig.~\ref{fig_lhc}) are available. To use them as the input one, the user 
should change the name of the corresponding file to the \verb*|RunInputHydjet|. 
The default parameters for \verb*|RunInputHydjetRHIC200| were
obtained by fitting RHIC data to various physical observables (see Section 6 
for the details). The default parameters for 
\verb*|RunInputHydjetLHC5500| represent our rough extrapolation from RHIC to LHC 
energy, and of course they may be varied freely by the user. 

The following input parameters should be specified by the user.  

\noindent 
\verb*|fNevnt| -- number of events to generate;\\
\verb*|fSqrtS| -- beam c.m.s. energy per nucleon pair in GeV; \\ 
\verb*|fAw| -- atomic weight of nuclei; \\ 
\verb*|fIfb|  -- flag of type of centrality generation ($=$\verb*|0|: impact 
parameter is fixed by \verb*|fBfix| value, $\ne$\verb*|0|: impact parameter is 
generated in each event between minimum (\verb*|fBmin|) and maximum (\verb*|fBmax|) 
values according to the distribution~(\ref{sigin_b})); \\
\verb*|fBmin| -- minimum impact parameter in units of nucleus radius $R_A$; \\
\verb*|fBmax| -- maximum impact parameter in units of nucleus radius $R_A$; \\
\verb*|fBfix| -- fixed impact parameter in units of nucleus radius $R_A$. 
  
The following input parameters may be changed by user from their default values. 

\noindent 
\verb*|fSeed| -- parameter to set the random number seed ($=$\verb*|0|: the current 
time is used to set the random generator seed, $\ne$\verb*|0|: the value 
\verb*|fSeed| is used to set the random generator seed and then the state of 
random number generator in PYTHIA is \verb*|MRPY(1)|=\verb*|fSeed|) 
(default: \verb*|0|).
 
Parameters for soft hydro-type part of the event.

\noindent 
\verb*|fT| -- chemical freeze-out temperature $T^{\rm ch}$ in GeV; \\
\verb*|fMuB| -- chemical baryon potential per unit charge $\widetilde{\mu_{B}}$ in 
GeV; \\
\verb*|fMuS| -- chemical strangeness potential per unit charge 
$\widetilde{\mu_{S}}$ in GeV;\\
\verb*|fMuI3| -- chemical isospin potential per unit charge $\widetilde{\mu_{Q}}$ 
in GeV;\\
\verb*|fTthFO| -- thermal freeze-out temperature $T^{\rm th}$ in GeV;\\ 
\verb*|fMu_th_pip| -- chemical potential of positively charged pions at thermal 
freeze-out $\mu_{\pi}^{\rm eff~th}$ in GeV; \\
\verb*|fTau| -- proper time at thermal freeze-out for central collisions
$\tau_f(b=0)$ in fm/$c$; \\
\verb*|fSigmaTau| -- duration of emission at thermal freeze-out for central 
collisions $\Delta \tau_f(b=0)$ in fm/$c$; \\
\verb*|fR| -- maximal transverse radius at thermal freeze-out for central 
collisions $R_f(b=0)$ in fm; \\
\verb*|fYlmax| -- maximal longitudinal flow rapidity at thermal 
freeze-out $\eta_{\rm max}$;\\
\verb*|fUmax| -- maximal transverse flow rapidity at thermal 
freeze-out for central collisions $\rho_u^{\rm max}(b=0)$;\\
\verb*|fDelta| -- momentum azimuthal anisotropy parameter at thermal freeze-out  
$\delta$ (for given event centrality class); \\
\verb*|fEpsilon| -- coordinate azimuthal anisotropy parameter at thermal 
freeze-out $\epsilon$ (for given event centrality class); \\
\verb*|fIfDeltaEpsilon|  -- flag to use calculated $\delta$ and $\epsilon$ 
($\le$\verb*|0|: specifed by user values for given event centrality class are 
taken, $>$\verb*|0|: calculated for given impact parameter according to the 
parameterization (\ref{v2-eps-delta2}) values are used in each event) 
(default: 0); \\
\verb*|fDecay| -- flag to switch on/off hadron decays ($<$\verb*|0|: decays ``off'',
$\ge$\verb*|0|: decays ``on'') (default: 0); \\
\verb*|fWeakDecay| -- flag to switch on/off weak hadron decays ($<$\verb*|0|: decays 
``off'', $\ge$\verb*|0|: decays ``on'') (default: 0); \\
\verb*|fEtaType| -- flag to specify longitudinal flow rapidity distribution 
($\le$\verb*|0|: uniform in the range $[$\verb*|-fYlmax,fYlmax|$]$, $>$\verb*|0|: 
Gaussian with the dispersion \verb*|fYlmax|) (default: \verb*|1|); \\
\verb*|fTMuType| -- flag to use calculated chemical freeze-out temperature, 
baryon potential and strangeness potential as a function of \verb*|fSqrtS|
($\le$\verb*|0|: specified by user values are taken, $>$\verb*|0|: calculated
according to the parameterization (\ref{Cleymans2}) values are used) 
(default: \verb*|0|); \\
\verb*|fCorrS| -- flag and value to include strangeness suppression factor 
$\gamma_s$ with \verb*|fCorrS| value ($0<$\verb*|fCorrS|$\le 1$, 
if \verb*|fCorrS|$\le0$ then it is calculated using its phenomenological 
dependence $\gamma_s = 1 - 0.386\exp{(-1.23T^{\rm ch}/\mu_{B})}$ 
from~\cite{Becattini:2005xt}) (default: \verb*|1.|).  

Parameters for treatment of hard multi-parton part of the event.

\verb*|fNhsel| -- flag to switch on/off jet and hydro-state production (\verb*|0|: jet 
production ``off'' and hydro ``on'', \verb*|1|: jet production ``on'' and jet quenching 
``off'' and hydro ``on'', \verb*|2|: jet production ``on'' and jet quenching ``on'' and
hydro ``on'', \verb*|3|: jet production ``on'' and jet quenching ``off'' and hydro 
``off'', \verb*|4|: jet production ``on'' and jet quenching ``on'' and hydro`` off'') 
(default: \verb*|2|); \\
\verb*|fIshad| -- flag to switch on/off impact parameter dependent nuclear 
shadowing for gluons and light sea quarks (u,d,s) (\verb*|0|: shadowing ``off'', 
\verb*|1|: shadowing ``on'' for \verb*|fAw|$=$\verb*|207.|, \verb*|197.|, \verb*|110.| 
or \verb*|40.|) (default: \verb*|1|); \\
\verb*|fPtmin| -- minimal transverse momentum transfer $p_T^{\rm min}$ of hard 
parton-parton scatterings in GeV/$c$ (the PYTHIA parameter \verb*|ckin(3)|=\verb*|fPtmin|) 

\noindent 
PYQUEN energy loss model parameters: \\   
\verb*|fT0| -- initial temperature $T_0$ (in GeV) of QGP for 
central Pb+Pb collisions at mid-rapidity (initial temperature for other 
centralities and atomic numbers will be calculated automatically) (allowed 
range is \verb*|0.2|$<$\verb*|fT0|$<$\verb*|2.|); \\
\verb*|fTau0| -- proper QGP formation time $\tau_0$ in fm/c 
(\verb*|0.01|$<$\verb*|fTau0|$<$\verb*|10|);\\
\verb*|fNf| -- number of active quark flavours $N_f$ in QGP 
(\verb*|fNf|=\verb*|0|, \verb*|1|, \verb*|2| or \verb*|3|); \\
\verb*|fIenglu| -- flag to fix type of in-medium partonic energy loss (\verb*|0|: 
radiative and collisional loss, \verb*|1|: radiative loss only, \verb*|2|: 
collisional loss only) (default: \verb*|0|); \\
\verb*|fIanglu| -- flag to fix type of angular distribution of in-medium emitted 
gluons (\verb*|0|: small-angular~(\ref{sar}), \verb*|1|: wide-angular, \verb*|2|: 
collinear) (default: 0). 

Note that if specified by user value of input parameter extends out of the 
allowed range, its default value is used in HYDJET++ run. 

A number of important PYTHIA parameters also may be changed/specified in 
\verb*|RunInputHydjet| file (they are not shown in Figs.~\ref{fig_inrhic} and 
\ref{fig_lhc}). The rest PYTHIA parameters can be changed (if it is necessary) 
using corresponding common blocks in the file \verb*|progs_fortran.f|.

\subsection{Output parameters and files}

The program output is directed to a ROOT file specified by the user in the command
line or by default to \verb*|RunOutput.root|. The output file contains a tree
named \verb*|td|, which keeps the entire event record including primary particles
and decay products with their coordinates and momenta information. Each decay
product contains the unique index of its parent particle so that the entire event history may be obtained.
Beside particle information, the output file contains also the following global
output parameters for each event. 

\noindent
\verb*|Bgen| -- generated value of impact parameter $b$ in units of nucleus radius 
$R_A$; \\
\verb*|Sigin| -- total inelastic NN cross section at given energy 
\verb*|fSqrtS| (in mb); \\ 
\verb*|Sigjet| -- hard scattering NN cross section at given \verb*|fSqrtS|,    
\verb*|fPtmin| (in mb); \\
\verb*|Ntot| -- generated value of total event multiplicity 
(\verb*|Ntot|=\verb*|Nhyd|+\verb*|Npyt|); \\ 
\verb*|Nhyd| -- generated multiplicity of ``soft'' hydro-induced particles; \\
\verb*|Npyt| -- generated multiplicity of ``hard'' jet-induced particles; \\ 
\verb*|Njet| -- generated number $N_{AA}^{\rm jet}$ of hard parton scatterings 
with $p_T>$fPtmin; \\
\verb*|Nbcol| -- mean number of binary NN sub-collisions $\overline{N_{\rm bin}}$
(\ref{nbcol}) at given \verb*|Bgen|; \\
\verb*|Npart| -- mean number of nucleons-participants $\overline{N_{\rm part}}$ 
(\ref{npart}) at given \verb*|Bgen|. 

The event output tree (\verb*|ROOT::TTree|) is organized as follows:

\begin{verbatim}
td->Branch("nev",&nev,"nev/I");        
        // event number 
td->Branch("Bgen",&Bgen,"Bgen/F"); 
        // generated impact parameter 
td->Branch("Sigin",&Sigin,"Sigin/F"); 
        // total inelastic NN cross section
td->Branch("Sigjet",&Sigjet,"Sigjet/F"); 
        // hard scattering NN cross section
td->Branch("Ntot",&Ntot,"Ntot/I");  
        //  total event multiplicity 
td->Branch("Nhyd",&Nhyd,"Nhyd/I");  
        //  multiplicity of hydro-induced particles
td->Branch("Npyt",&Npyt,"Npyt/I");  
        //  multiplicity of jet-induced particles
td->Branch("Njet",&Njet,"Njet/I"); 
        // number of hard parton-parton scatterings 
td->Branch("Nbcol",&Nbcol,"Nbcol/I");  
        // mean number of NN sub-collisions
td->Branch("Npart",&Npart,"Npart/I"); 
        // mean number of nucleon-participants
td->Branch("Px",&Px[0],"Px[npart]/F"); 
        // x-component of the momentum, in GeV/c
td->Branch("Py",&Py[0],"Py[npart]/F"); 
        // y-component of the momentum, in GeV/c
td->Branch("Pz",&Pz[0],"Pz[npart]/F"); 
        // z-component of the momentum, in GeV/c
td->Branch("E",&E[0],"E[npart]/F");    
        // energy, in GeV 
td->Branch("X",&X[0],"X[npart]/F");    
        // x-coordinate at emission point, in fm
td->Branch("Y",&Y[0],"Y[npart]/F");    
        // y-coordinate at emission point, in fm
td->Branch("Z",&Z[0],"Z[npart]/F");    
        // z-coordinate at emission point, in fm
td->Branch("T",&T[0],"T[npart]/F");    
        // proper time of particle emission, in fm/c 
td->Branch("pdg",&pdg[0],"pdg[npart]/I");    
        // Geant particle code
td->Branch("Mpdg",&Mpdg[0],"Mpdg[npart]/I"); 
        // Geant mother code (-1 for primordial)
td->Branch("type",&type[0],"type[npart]/I");
        // particle origin (=0: hydro, >0: jet) 
td->Branch("Index",&Index[0],"Index[Ntot]/I");
        // unique zero based index of the particle
td->Branch("MotherIndex",&MotherIndex[0],"MotherIndex[Ntot]/I");
        // index of the mother particle (-1 if its a primary particle)
td->Branch("NDaughters",&NDaughters[0],"NDaughters[Ntot]/I");
        // number of daughter particles
td->Branch("Daughter1Index",&Daughter1Index[0],"Daughter1Index[Ntot]/I");
        // index of the first daughter (-1 if it does not exist)
td->Branch("Daughter2Index",&Daughter2Index[0],"Daughter2Index[Ntot]/I");
        // index of the second daughter
td->Branch("Daughter3Index",&Daughter3Index[0],"Daughter3Index[Ntot]/I");
        // index of the third daughter
\end{verbatim} 

The possibility to create event output written directly to histograms (in 
according to user's specification in the file \verb*|RunHadronSourceHisto.cxx|) is 
envisaged. The histogram output is directed by default to the file \verb*|RunOutputHisto.root|
or to a file specified explicitly by the user in the command line.
  
\subsection{Organization of the Fortran package}

The hard, multi-jet part of HYDJET++ event is identical to the hard part of 
Fortran-written HYDJET~\cite{Lokhtin:2005px,hydjet} (version 1.5). The three 
Fortran files are placed in the directory PYQUEN. 

\verb*|pythia-6.4.11.f| -- PYTHIA event generator to simulate hard NN 
sub-\-col\-lisions \cite{pythia} (is used if \verb*|fNhsel|=1, 2, 3 or 4).

\verb*|pyquen1_5.f| -- PYQUEN event generator to modify PYTHIA-produced hard 
events according to the corresponding partonic energy loss 
model~\cite{Lokhtin:2005px,pyquen} (is used if \verb*|fNhsel|=2 or 4), and to 
generate the spatial vertex of a jet production. 

\verb*|progs_fortran.f| -- main Fortran routine, which contains the following main
subroutines and functions (the auxiliary subroutines and functions are not 
listed here). 

$\bullet$ \verb*|subroutine hyinit(fSqrtS,fAw,fIfb,fBmin,fBmax,fBfix)| -- 
initializes PYTHIA, calculates cross sections \verb*|Sigin| and \verb*|Sigjet|, 
tabulates $T_A(b)$, $T_{AA}(b)$.  

$\bullet$ \verb*|subroutine hyevnt| -- in each event generates the impact parameter 
value \verb*|Bgen| (if \verb*|fIfb| $\ne 0$), calculates numbers \verb*|Nbcol| and 
\verb*|Npart|, generates number of ``successful'' hard NN sub-collisions 
\verb*|Njet|, and calls \verb*|hyhard| subroutine. 

$\bullet$ \verb*|subroutine hyhard| -- generates multi-parton production in 
\verb*|Njet| hard NN sub-collisions (with jet quenching if \verb*|fNhsel|=2 or 4, and 
with nuclear shadowing if \verb*|fIshad|=1), performs hadronization, and writes 
the final event output in the array \verb*|hyjets|.

$\bullet$ \verb*|subroutine ggshad(inucl,x,q2,b,res,taf)| -- adapts the parameterization
of ratio of nuclear to nucleon parton distribution function \verb*|res| (\verb*|res(1)| 
for gluons and \verb*|res(2)| for sea quarks) in a given nucleus \verb*|inucl|, Bjorken 
\verb*|x| in the parton distribution, square of transverse momentum transfer in the hard 
scattering \verb*|q2|, and transverse position of jet production vertex relative to the 
nucleus center \verb*|b| (\verb*|taf| is the parametrized nuclear thickness function). 

$\bullet$ \verb*|function shad1(kfh,xbj,q2,r)| -- in each event calculates the ratio of 
nuclear to nucleon parton distribution function (normalized by the atomic number) for the  
given parton code \verb*|kfh|, Bjorken x in a parton distribution \verb*|xbj|, square 
of transverse momentum transfer in the hard scattering \verb*|q2|, and transverse 
position of jet production vertex relative to the the nucleus center \verb*|r| 
(\verb*|call ggshad| is performed).
 
\subsection{Organization of the C++ package}

The organization of the C++ package in HYDJET++ is similar to the FAST MC framework 
presented in~\cite{Amelin:2006qe,Amelin:2007ic}. The C++ source files are placed in 
the main directory. 

{\bf The main modules of the C++ package}.    

$\bullet$ \verb*|RunHadronSource.cxx|, \verb*|RunHadronSourceHisto.cxx| -- contains 
the \verb*|main| program. It handles the \verb*|InitialStateHydjet| class instance by 
calling the \verb*|ReadParams()|, \verb*|MultiIni()|, \verb*|Initialize()| and 
\verb*|Evolve()| member functions. The \verb*|main| program also creates the 
output ROOT tree and fills it with event information.

$\bullet$ \verb*|InitialState.cxx|, \verb*|InitialState.h| -- contains virtual 
class \verb*|InitialState|. The \verb*|InitialState| class is responsible for 
initialization of the PDG particle database (\verb*|DatabasePDG|). The function 
\verb*|Evolve()| organizes the particle decays and updates the coordinate-time 
information on the particle list.

$\bullet$ \verb*|InitialStateHydjet.cxx|, \verb*|InitialStateHydjet.h| -- contains the 
definition of the class \verb*|InitialStateHydjet|. This class inherits the upper 
class \verb*|InitialState|. The member function \verb*|ReadParams()| reads the parameters 
from the input file once per run. The function \verb*|MultiIni()| calculates the total 
particle specie multiplicities once per run. The \verb*|Initialize()| function 
generates the particles in the initial fireball. The \verb*|Evolve()| function, the  
heritage of the class \verb*|InitialState|, performs the decay of unstable 
particles.
 
$\bullet$ \verb*|HadronDecayer.cxx|, \verb*|HadronDecayer.h| -- contains two functions called 
by the function \verb*|InitialState::Evolve()| to decay the resonances: \verb*|Decay()| and 
\verb*|GetDecayTime()|.

{\bf The service modules of the C++ package}. 

$\bullet$ \verb*|Particle.cxx|, \verb*|Particle.h| -- contains the \verb*|Particle| class which 
handles all the track information (PDG properties, momenta, coordinate, indexes of parent
particles and decay products). 

$\bullet$ \verb*|ParticlePDG.cxx|, \verb*|ParticlePDG.h| -- contains the definition of the class  
\verb*|ParticlePDG| which stores all the PDG properties of a particle specie. This class 
is used by the \verb*|Particle| class and by the \verb*|DatabasePDG| to organize the information.
 
$\bullet$ \verb*|DatabasePDG.cxx|, \verb*|DatabasePDG.h| -- contains the definition of the class 
\verb*|DatabasePDG|. This class reads and handles all the particle specie definitions 
found in \verb*|particles.data| and all the decay channels found in \verb*|tabledecay.txt|.

$\bullet$ \verb*|DecayChannel.cxx|, \verb*|DecayChannel.h| -- contains the \verb*|DecayChannel| 
class definition. This class stores the information for a single decay channel. A 
\verb*|ParticlePDG| instance uses  \verb*|DecayChannel| objects to store informations regarding 
decay channels.

$\bullet$ \verb*|GrandCanonical.cxx|, \verb*|GrandCanonical.h| -- contains the method to  
calculate the densities of energy, baryon and electric charge and particle 
numbers, within the grand canonical approach by means of the temperature and 
chemical potentials. 

$\bullet$ \verb*|StrangeDensity.cxx|, \verb*|StrangeDensity.h| -- contains the method to 
calculate the stran\-ge\-ness density within the grand canonical description.

$\bullet$ \verb*|HankelFunction.cxx|, \verb*|HankelFunction.h| -- computes modified 
Hankel function of zero, first and second orders. 

$\bullet$ \verb*|StrangePotential.cxx|,\verb*|StrangePotential.h| -- contains the method 
to calculate strange potential (if \verb*|fTMuType =1|) from the initial strange density at 
given temperature and baryon potential.

$\bullet$ \verb*|EquationSolver.cxx|, \verb*|EquationSolver.h|  -- is used by 
\verb*|StrangePotential| class to calculate strangeness potential. 
 
$\bullet$ \verb*|RandArrayFunction.cxx|, \verb*|RandArrayFunction.h| -- defines several 
methods for shooting generally distributed random values, given a user-defined 
probability distribution function.

$\bullet$ \verb*|UKUtility.cxx|, \verb*|UKUtility.h| -- contains the method \verb*|IsotropicR3| 
to generate the three-vector uniformily distributed on spherical surface, and the method 
\verb*|MomAntiMom| to calculate kinematical variables for two-body decays.  
  
$\bullet$ \verb*|MathUtil.h| -- contains some useful constant determinations. 

$\bullet$ \verb*|HYJET_COMMONS.h| -- contains the list of Fortran common blocks used by C++ 
package.

\section{Installation instructions}

HYDJET++ package is available via Internet. It can be downloaded from the 
corresponding web-page~\cite{hydjet++} as the archive \verb*|HYDJET++.ZIP|, which 
contains the following files and directories: \\
$\bullet$ the Makefile; \\
$\bullet$ the input files (\verb*|RunInputHydjet|, \verb*|RunInputHydjetRHIC200|,\\ 
\verb*|RunInputHydjetLHC5500|); \\
$\bullet$ \verb*|HYDJET| directory with \verb*|inc| (\verb*|*.h| files), \verb*|src| 
(\verb*|*.cxx| files) and SHARE data files (\verb*|particle.data|,
\verb*|tabledecay.txt|); \\
$\bullet$ \verb*|PYQUEN| directory with \verb*|*.f| files; \\ 
$\bullet$ \verb*|DOC| directory (with short description and update notes for new
versions); \\
$\bullet$ \verb*|RootMacros| directory (the examples allowing one to obtain 
some physical results with HYDJET++, detailed comments are in the macros).
  
In order to run HYDJET++ on Linux one needs:\\
1) C++ and Fortran compilers; \\
2) ROOT libraries and include files.\\
The main program is in the files \verb*|RunHadronSource.cxx| (for ROOT tree 
output) or \verb*|RunHadronSourceHisto.cxx| (for ROOT histogram output). To 
compile the package one needs to use the following commands in the main \\ 
HYDJET directory: \\
\verb*|make| (for ROOT tree output), or \\ 
\verb*|make -f Makefile_HISTO| (for ROOT histogram output). \\ 
Then the executable file \verb*|HYDJET| (or \verb*|HYDJET_HISTO|) is created in 
the same directory. Once the program is compiled, one can use the executables above
to start a simulation run. The input file necessary to run HYDJET++ must always be 
named \verb*|RunInputHydjet|, so after preparing this file according to the examples
included in the package one can start the simulation with the command line: 
\verb*|./HYDJET| (or \verb*|./HYDJET_HISTO|). If an output file is not 
specified then the event records will be automatically directed to the file  
\verb*|RunOutput.root| (or \verb*|RunOutputHisto.root|).
         
\section{Validation of HYDJET++ with experimental RHIC data}  

It was demonstrated in previous papers that FAST MC model can describe well 
the bulk properties of hadronic state created in Au+Au collisions at RHIC at
$\sqrt{s}=200 A$ GeV (such as particle number ratios, low-p$_T$ spectra, 
elliptic flow coefficients $v_2(p_T, b)$, femtoscopic correlations in 
central collisions)~\cite{Amelin:2006qe,Amelin:2007ic}, while HYDJET model is 
capable of reproducing the main features of jet quenching pattern at RHIC 
(high-$p_T$ hadron spectra and the suppression of azimuthal back-to-back 
correlations)~\cite{Lokhtin:2005px}. Since soft and hard hadronic states in
HYDJET++ are simulated independently, a good description of hadroproduction at 
RHIC in a wide kinematic range can be achieved, moreover a number of 
improvements in FAST MC and HYDJET have been done as compared to earlier 
versions. A number of input parameters of the model can be fixed from fitting 
the RHIC data to various physical observables (listed in Fig.~\ref{fig_inrhic}). 

\begin{enumerate}  

\item {\bf Ratio of hadron abundances.} It is well known that the particle
abundances in heavy ion collisions in a wide energy range can be reasonable well
described within statistical models based on the assumption that the produced
hadronic matter reaches thermal and chemical equilibrium. The thermodynamical
parameters $\widetilde{\mu_{B}}=0.0285$ GeV, $\widetilde{\mu_{S}}=0.007$ GeV, 
$\widetilde{\mu_{Q}}=-0.001$, the strangeness suppression factor $\gamma_s=1$, 
and the chemical freeze-out temperature  $T^{\rm ch}=0.165$ GeV have been fixed 
in~\cite{Amelin:2006qe} from fitting the RHIC data to  
various particle ratios near mid-rapidity in central Au+Au collisions at 
$\sqrt{s}=200 A$ GeV ($\pi^-/\pi^+$, $\bar{p}/\pi^-$, $K^-/K^+$, $K^-/\pi^-$, 
$\bar{p}/p$, $\bar{\Lambda}/\Lambda$, $\bar{\Lambda}/\Lambda$, $\bar{\Xi}/\Xi$, 
$\phi/K^-$, $\Lambda/p$, $\Xi^-/\pi^-$). 

\item {\bf Low-$p_T$ hadron spectra.} Transverse momentum $p_T$ and transverse 
mass $m_T$ hadron spectra ($\pi^+$, $K^+$ and $p$ with $m_T<0.7$ GeV/$c^2$) near  
mid-rapidity at different centralities of Au+Au collisions at $\sqrt{s}=200 A$ 
GeV were analyzed in~\cite{Amelin:2007ic}. The slopes of these
spectra allow the thermal freeze-out temperature  $T^{\rm th}=0.1$ GeV and 
the maximal transverse flow rapidity in central collisions 
$\rho_u^{\rm max}(b=0)=1.1$ to be fixed.  

\item {\bf Femtoscopic correlations.} Because of the effects of quantum 
statistics and final state interactions, the momentum (HBT) correlation functions 
of two or more particles at small relative momenta in 
their c.m.s. are sensitive to the space-time characteristics of 
the production process on the level of $fm$. The space-time parameters of thermal 
freeze-out region in central Au+Au collisions at $\sqrt{s}=200 A$ GeV have been 
fixed in~\cite{Amelin:2007ic} by means of fitting the three-dimensional correlation 
functions measured for $\pi^+\pi^+$ pairs and extracting the correlation radii 
$R_{\rm side}$, $R_{\rm out}$ and $R_{\rm long}$: $\tau_f(b=0)=8$ fm/$c$, 
$\Delta \tau_f(b=0)=2$ fm/$c$, $R_f(b=0)=10$ fm. 

\item {\bf Pseudorapidity hadron spectra.} The PHOBOS data on $\eta$-spectra of 
charged hadrons~\cite{Back:2002wb} at different centralities of Au+Au collisions 
at $\sqrt{s}=200 A$ GeV have been analyzed to fix the particle densities in the 
mid-rapidity region and the maximum longitudinal flow rapidity 
$\eta_{\rm max}=3.3$ (Fig. \ref{fig_eta_all}). Since mean ``soft'' and ``hard''
hadron multiplicities depend on the centrality in different ways (they are 
roughly proportional to $\overline{N_{\rm part}(b)}$ and 
$\overline{N_{\rm bin}(b)}$ respectively), the relative contribution of soft 
and hard parts to the total event multiplicity can be fixed through the 
centrality dependence of $dN/d\eta$. The corresponding contributions from 
hydro- and jet-parts are determined by the input parameters 
$\mu_{\pi}^{\rm eff~th}=0.053$ GeV and $p_T^{\rm min}=3.4$ GeV/$c$ respectively.

\begin{figure}
\begin{center}
\includegraphics[width=14cm]{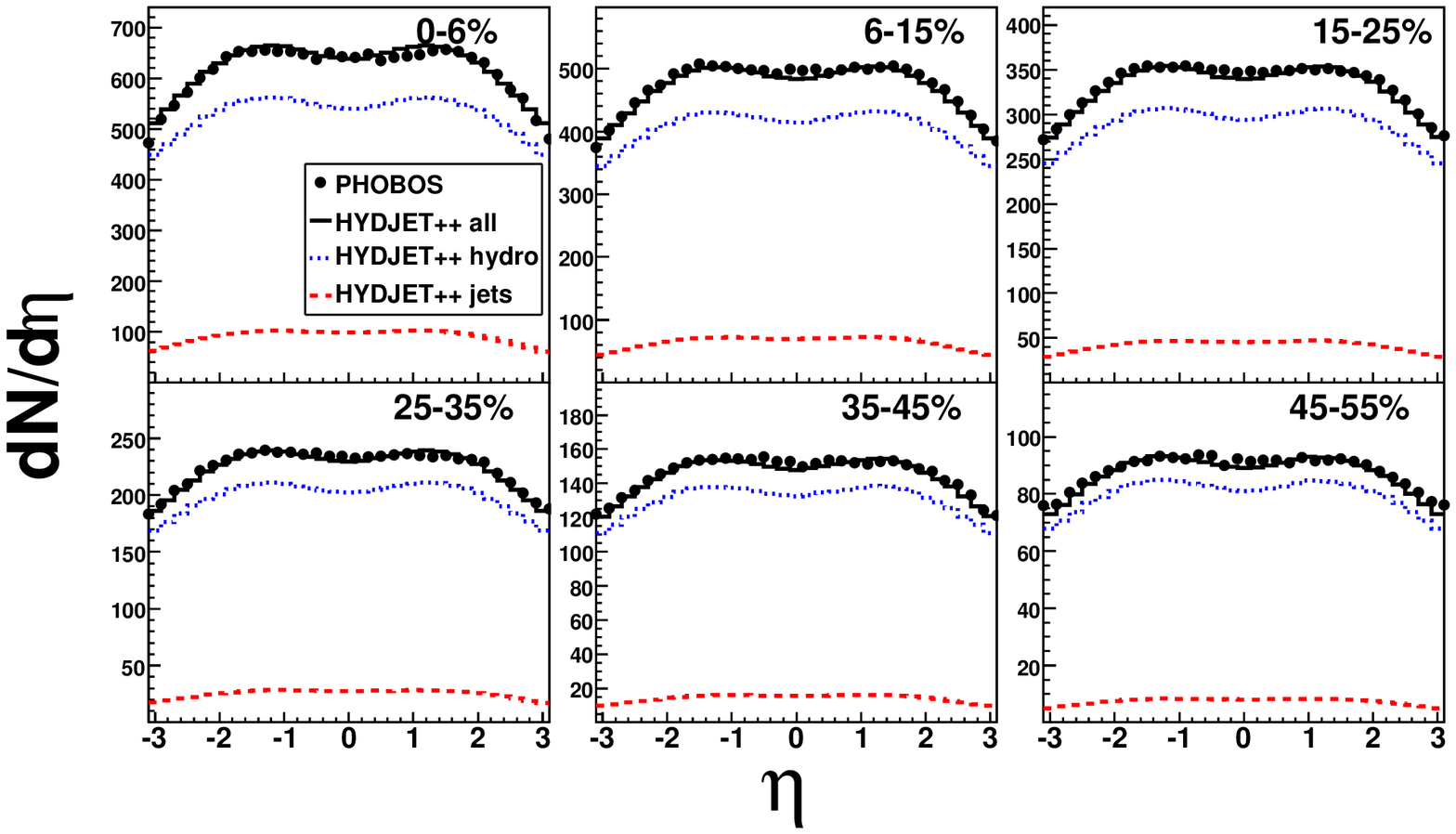}
\end{center}
\caption{The pseudorapidity distribution of charged hadrons in Au+Au 
collisions at $\sqrt{s}=200 A$ GeV for six centrality sets. The points are 
PHOBOS data~\cite{Back:2002wb}, histograms are the HYDJET++ calculations 
(solid -- total, dotted -- jet part, dashed -- hydro part).}
\label{fig_eta_all}
\end{figure}

\begin{figure}
\begin{center}
\includegraphics[width=14cm]{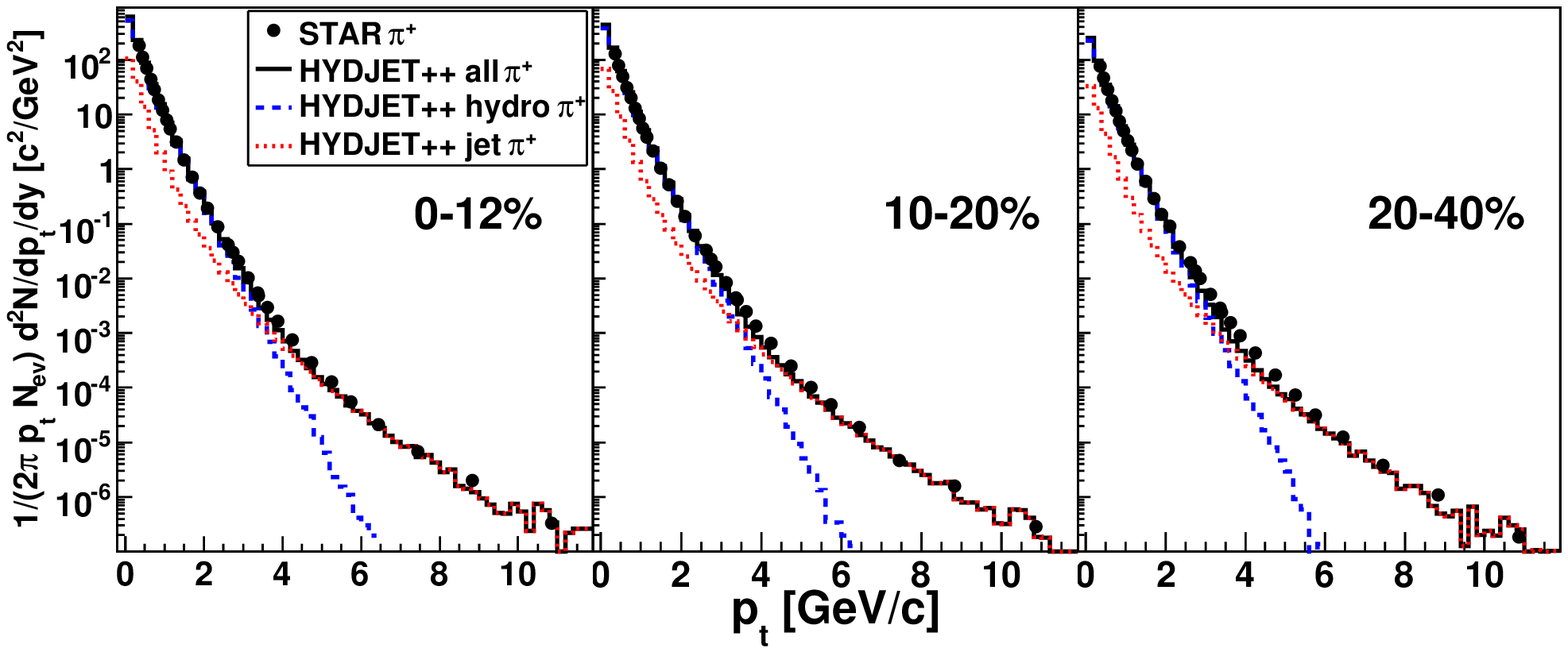}
\end{center}
\caption{The transverse momentum distribution of positively charged pions in 
Au+Au collisions at $\sqrt{s}=200 A$ GeV for three centrality sets. The points 
are STAR data~\cite{Abelev:2006jr}, histograms are the HYDJET++ calculations 
(solid -- total, dotted -- jet part, dashed -- hydro part).}
\label{fig_hpt_all}
\end{figure}

\begin{figure}
\begin{center}
\includegraphics[width=14cm]{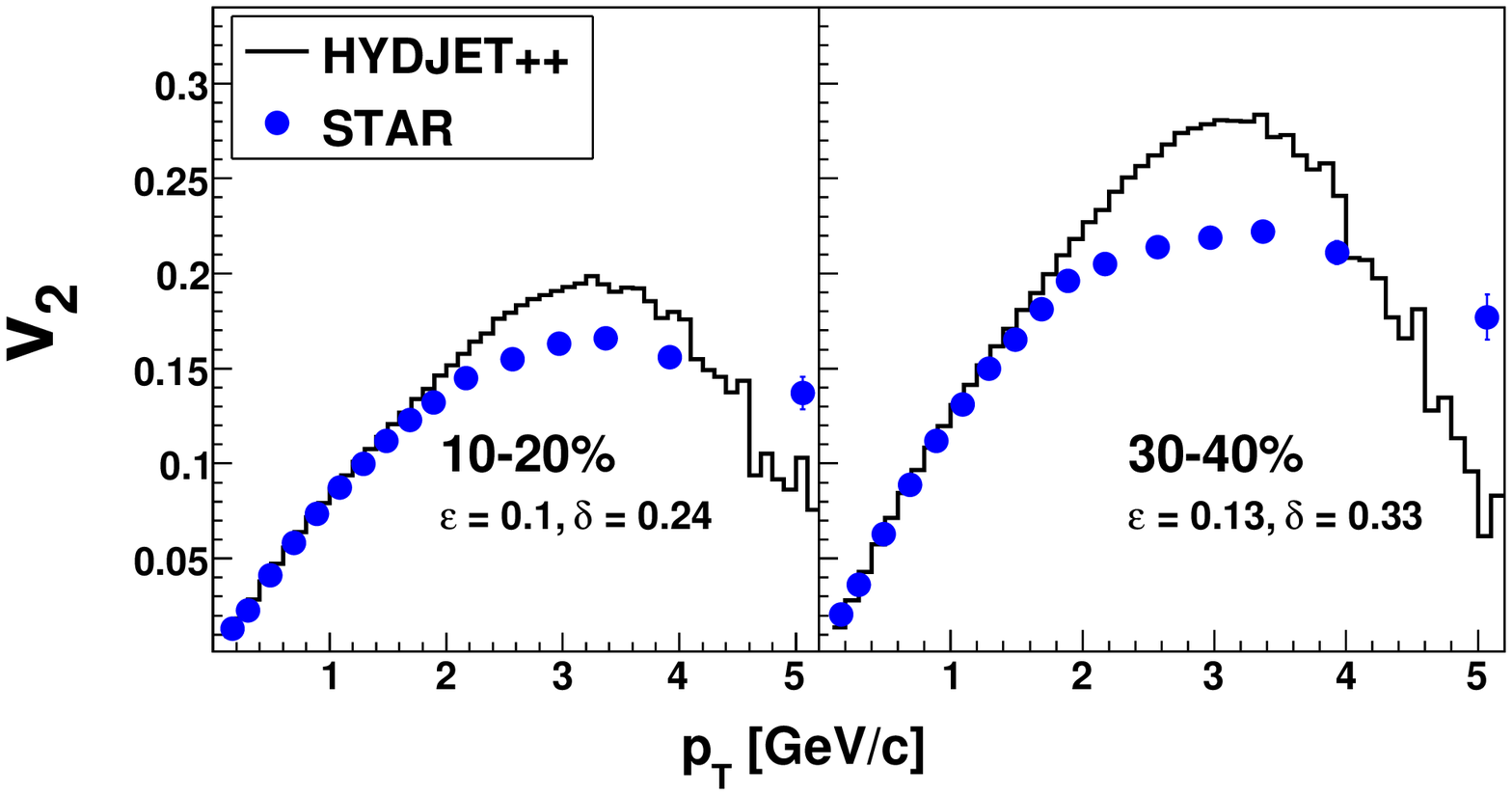}
\end{center}
\caption{The transverse momentum dependence of the elliptic flow  
coefficient $v_2$ of charged hadrons in Au+Au collisions at $\sqrt{s}=200 A$ 
GeV for two centrality sets. The points are STAR data~\cite{Adams:2004bi}, 
histograms are the HYDJET++ calculations.}
\label{fig_v2_all}
\end{figure}

\item {\bf High-$p_T$ hadron spectra.} High transverse momentum hadron 
spectra ($p_T \ga 2-4$ GeV/$c$) are sensitive to parton production and 
jet quenching effects. Thus fitting the measured high-$p_T$ tail allows 
the extraction of PYQUEN energy loss model parameters. We assume 
the QGP formation time $\tau_0=0.4$ fm/$c$ and the number of active quark 
flavours $N_f=2$. Then the reasonable 
fit of STAR data on high-$p_T$ spectra of charged pions at different 
centralities of Au+Au collisions at $\sqrt{s}=200 A$ GeV~\cite{Abelev:2006jr} is
obtained with the initial QGP temperature $T_0=0.3$ GeV (Fig. \ref{fig_hpt_all}).  

\item {\bf Elliptic flow.} The elliptic flow coefficient $v_2$ (which is
determined as the second-order Fourier coefficient in the hadron distribution
over the azimuthal angle $\varphi$ relative to the reaction plane angle
$\psi_R$, so that $v_2 \equiv \left< \cos{2(\varphi-\psi_R)} \right>$) is an
important signature of the physics dynamics at early stages of non-central 
heavy ion collisions. According to the typical hydrodynamic scenario, the
values $v_2(p_T)$ at low-$p_T$ ($\la 2$ GeV/$c$) are determined mainly by the 
internal pressure gradients of an expanding fireball during the initial high 
density phase of the reaction (and it is sensitive to the momentum and azimuthal 
anisotropy parameters $\delta$ and $\epsilon$ in the frameworks of HYDJET++), while 
elliptic flow at high-$p_T$ is generated in HYDJET++ (as well as in other jet
quenching models) due to the partonic energy loss in an azimuthally asymmetric 
volume of QGP. Figure \ref{fig_v2_all} shows the measured by the STAR 
Collaboration transverse momentum dependence of the elliptic flow coefficient 
$v_2$ of charged hadrons in Au+Au collisions at $\sqrt{s}=200 A$ GeV for two  
centrality sets~\cite{Adams:2004bi}. The values of $\delta$ and $\epsilon$ 
estimated for each centrality are written on the plots. Note that the choice 
of these parameters does not affect any azimuthally integrated physics 
observables (such as hadron multiplicities, $\eta$- and $p_T$-spectra, etc.), 
but only their differential azimuthal dependences. 

\end{enumerate}

\section{Test run description} 

The set of macros for comparison of various physics observables with the 
experimental data obtained for different centralities of Au+Au collisions at 
$\sqrt{s}=200 A$ GeV are in the directory \verb*|RootMacros|. 

$\bullet$ \verb*|fig_eta_Phobos.C|, \verb*|fig_eta_Phobos_read.C| -- the charged 
hadron pseudorapidity spectra are generated (Fig. \ref{fig_eta_all}). To start 
the generation, the user should type: \verb*|root -l fig_eta_Phobos.C|, or 
\verb*|root -l fig_eta_Phobos_read.C|. \\ In the latter case, the number of events 
should be specified in the macro \\ \verb*|fig_eta_Phobos_read.C|. The PHOBOS data
\cite{Back:2002wb} are in the directory \\ \verb*|eta_PHOBOS|.  

$\bullet$ \verb*|fig_PTH_STAR.C|, \verb*|fig_PTH_STAR_read.C| -- the positive pion 
$p_T$-spectra are generated (Fig. \ref{fig_hpt_all}). To start the generation, 
the user should type: \\
\verb*|root -l fig_PTH_STAR.C|, or \verb*|root -l fig_PTH_STAR_read.C|.
In the latter case, the number of events should be specified in the macro \\
\verb*|fig_PTH_STAR_read.C|. The STAR data \cite{Abelev:2006jr} are in the 
directory \verb*|HPT_STAR|. 

$\bullet$ \verb*|fig_v2_STAR.C|, \verb*|fig_v2_STAR_read.C| -- the elliptic flow 
coefficients $v_2(p_T)$ of charged hadrons are generated (Fig. \ref{fig_v2_all}). 
To start the generation, the  user should type: \verb*|root -l fig_v2_STAR.C|, 
or \verb*|root -l fig_v2_STAR_read.C|. In the latter case, the number of events 
should be specified in the macro \\ \verb*|fig_v2_test_read.C|. The STAR data 
\cite{Adams:2004bi} are in the directory \verb*|v2_STAR|. 

\section{Conclusion}

Ongoing and future experimental studies of relativistic heavy ion collisions 
require the development of new Monte-Carlo event generators and improvement of 
existing ones. The main advantage of MC technique for the simulation of multiple 
hadroproduction is that it allows visual comparison of theory and   
data, including if necessary the detailed detector acceptances, responses and 
resolutions. The realistic MC event generator should include a maximum possible 
number of observable physical effects which are important to  
determine the event topology: from bulk properties of soft 
hadroproduction (domain of low transverse momenta $p_T$) to hard multi-parton 
production in hot QCD-matter, which reveals itself in spectra of 
high-$p_T$ particles and hadronic jets. 

The HYDJET++ event generator has been developed to simulate relativistic heavy ion 
AA collisions considered as a superposition of the soft, hydro-type state and the 
hard state resulting from multi-parton fragmentation. It includes detailed treatment 
of soft hadroproduction (collective flow phenomena and resonance decays) as well as 
hard parton production, and takes into account the known medium effects (as jet
quenching and nuclear shadowing). The main program is written in the 
object-oriented C++ language under the ROOT environment. The hard part of 
HYDJET++ is identical to the hard part of 
Fortran-written HYDJET generator and it is included in the generator structure as 
a separate directory. The soft part of HYDJET++ event is the ``thermal'' 
hadronic state generated on the chemical and thermal freeze-out hypersurfaces 
obtained from the parameterization of relativistic hydrodynamics with preset  
freeze-out conditions (the adapted C+ code FAST MC). HYDJET++ is capable of 
simultaneous reproducing the various hadronic observables measured in heavy ion  
collisions at RHIC for different centrality sets in a wide kinematic 
range: ratio of hadron yields, pseudorapidity and transverse momentum 
spectra (for both low- and high-$p_T$ domains), elliptic flow coefficients 
$v_2(p_T)$, femtoscopic correlations. Although the HYDJET++ generator is 
optimized for RHIC and LHC energies, in practice it can be also used for 
studying of multi-particle production in a wider energy range down to 
$\sqrt{s} \sim 10$ GeV per nucleon pair at other heavy ion experimental facilities.
 
\section{Acknowledgements}

Discussions with L.V.~Bravina, D.~d'Enterria, G.H.~Eyyubova, A.I.~Demianov, 
I.M.~Dremin, A.B.~Kaidalov, Iu.A.~Kar\-pen\-ko, O.L.~Kodolova, V.L.~Korotkikh, 
R.~Lednicky, C.~Loizides, C.~Mironov, S.V.~Molodtsov, A.~Morsch, 
T.A.~Pocheptsov, C.~Roland, L.I.~Sary\-che\-va, Yu.M.~Sinyukov, T.~Sjostrand, 
C.Yu.~Tep\-lov, I.N.~Vardanyan, G.~Veres, I.~Vitev, B.~Wyslouch, Y.~Yilmaz, 
E.E.~Zabrodin, B.G.~Zakharov and G.M.~Zinovjev are gratefully acknowledged. 
This work was supported by Russian Foundation for Basic Research (grants No 
08-02-91001 and No 08-02-92496), Grants of President of Russian Federation 
(No 107.2008.2 and 1456.2008.2) and Grant of INTAS No 05-103-7484.

\end{document}